%% file: manuscript.tex
\documentclass[useAMS,usenatbib]{mnras}

\newcommand{\msun}{\ensuremath{\mathrm{M_{\odot}}}}
\newcommand{\e}[1]{\times 10^{#1}}

\input{packageIncludes}

\input{figureMacros}

\DeclareSIUnit[number-unit-product = \;]
\solarmass{\msun}

\DeclareSIUnit[number-unit-product = \;]
\dyn{dyn}

\DeclareSIUnit[number-unit-product = \;]
\erg{erg}

\makeatletter
\g@addto@macro\@floatboxreset\centering
\makeatother

\hypersetup{
    colorlinks,
    citecolor=blue,
    filecolor=black,
    linkcolor=blue,
    urlcolor=black
}

\title[Binary white dwarf mergers]{No double detonations but core carbon ignitions in high-resolution, grid-based simulations of binary white dwarf mergers}
\author[Fenn, Plewa, \& Gawryszczak]{D. Fenn$^{1}$\thanks{E-mail:df11c@my.fsu.edu}, 
T. Plewa$^{1}$\thanks{E-mail:tplewa@fsu.edu}, 
A. Gawryszczak$^{2}$\thanks{E-mail:gawrysz@camk.edu.pl}\\
$^{1}$Department of Scientific Computing, Florida State University, Tallahassee, FL 32306, U.S.A.\\
$^{2}$Nicolaus Copernicus Astronomical Center, Bartycka 18, Warsaw PL-00-716, Poland}
\begin{document}

\date{DATES HERE}

\pagerange{\pageref{firstpage}--\pageref{lastpage}} \pubyear{2016}

\maketitle

\label{firstpage}

\begin{abstract}
\input{abstract}
\end{abstract}

\begin{keywords}
accretion --- stars:white dwarfs --- binaries: close --- hydrodynamics: instabilities, shock waves --- supernovae:general --- nuclear reactions
\end{keywords}

\section{Introduction}\label{s:intro}
\input{introduction}

\section{Models and methods}\label{s:methods}
\input{modelsMethods}
\section{Results}\label{s:results}
\input{results}

\section{Discussion}\label{s:discussion}
\input{discussion}
\section{Summary}\label{s:summary}
\input{summary}
\section*{Acknowledgments}
We would like to thank the reviewer for comments that improved the clarity of the presentation. D.F. was supported by the DoD SMART Scholarship. T.P. was partially supported by the NSF grant
AST-1109113 and the DOE grant DE-FG52-09NA29548. This research used
resources of the National Energy Research Scientific Computing Center,
which is supported by the Office of Science of the U.S.\ Department of
Energy under Contract No.\ DE-AC02-05CH11231, and the Air Force Research Laboratory. The software used in
this work was in part developed by the DOE Flash Center at the
University of Chicago. The data analysis was performed in part using the yt Project \citep{turk+2011}, and VisIt \citep{HPV:VisIt}. This research has made use of NASA's Astrophysics Data System Bibliographic Services.
\appendix
\input{appendix}
\bibliographystyle{mnras}
\bibliography{manuscript}
\bsp
\label{lastpage}
\end{document}

%% file: packageIncludes.tex
\usepackage{amssymb}
\usepackage{fp}
\usepackage{xfrac}
\usepackage{amsmath}
\usepackage{hyperref}
\usepackage{booktabs}
\usepackage{cancel}
\usepackage{colortbl}
\usepackage{csquotes}
\usepackage{datatool}
\usepackage{helvet}
\usepackage{mathpazo}
\usepackage{listings}
\usepackage{multirow}
\usepackage{array}
\usepackage{float}

\usepackage{siunitx}

\usepackage{xparse}
\usepackage{etoolbox}
\usepackage{isotope}
\usepackage{ctable}
\usepackage{url}


%% file: figureMacros.tex
\newlength\figureheight 
\newlength\figurewidth 
\newcommand{\adjustThreePanels}{
    \setlength\figureheight{6cm}\ignorespaces 
    \setlength\figurewidth{6cm}\ignorespaces
    \setlength{\tabcolsep}{1pt}\ignorespaces
}


%% file: abstract.tex
We study the violent phase of the merger of massive binary white dwarf systems. Our aim is to characterize the conditions for explosive burning to occur, and identify a possible explosion mechanism of Type Ia supernovae. The primary components of our model systems are carbon-oxygen (C/O) white dwarfs, while the secondaries are made either of C/O or of pure helium. We account for tidal effects in the initial conditions in a self-consistent way, and consider initially well-separated systems with slow inspiral rates. We study the merger evolution using an adaptive mesh refinement, reactive, Eulerian code in three dimensions, assuming symmetry across the orbital plane. We use a co-rotating reference frame to minimize the effects of numerical diffusion, and solve for self-gravity using a multi-grid approach. We find a novel detonation mechanism in C/O mergers with massive primaries. Here the detonation occurs in the primary's core and relies on the combined action of tidal heating, accretion heating, and self-heating due to nuclear burning. The exploding structure is compositionally stratified, with a reverse shock formed at the surface of the dense ejecta. The existence of such a shock has not been reported elsewhere. The explosion energy (\SI{1.6e51}{erg}) and \isotope[56]{Ni} mass (\SI{0.86}{\solarmass}) are consistent with a SN Ia at the bright end of the luminosity distribution, with an approximated decline rate of $\Delta m_{15} (\mathrm{B}) \approx 0.99$. Our study does not support double-detonation scenarios in the case of a system with a \SI{0.6}{\solarmass} helium secondary and a \SI{0.9}{\solarmass} primary. Although the accreted helium detonates, it fails to ignite carbon at the base of the boundary layer or in the primary's core.

%% file: introduction.tex
Though decades of research into Type Ia supernovae (SNe Ia) have provided a significant quantity of information, critical details concerning the origins of these thermonuclear events are still missing. In particular, the nature of their progenitors and the mechanism of explosion, the two critical ingredients of the initial conditions, remain unkown. To complicate the picture, observations provide evidence for substantial diversity of SNe Ia, raising the possibility that both the progenitors and explosion mechanism differ among thermonuclear supernovae.

Despite the aforementioned lack of detailed knowledge, it is commonly agreed that these events are products of white dwarf explosions \citep[see, e.g.,][]{nugent+2011}. Two major theories dominate the landscape of hypothesized SN Ia progenitor scenarios. The first is the single degenerate (SD) scenario, in which a white dwarf star exists in a binary system with another, non-degenerate companion \citep{whelan+73, nomoto82b}. The second theory--namely, the double degenerate (DD) scenario, considers a binary system of white dwarf stars which, through radiation of gravitational waves, gradually loses angular momentum, resulting in a slow inspiral of the stellar components \citep{iben+84, webbink+84}. Although the conditions for ignition in these scenarios most likely differ, the supernova in both cases is powered by thermonuclear burning of degenerate material. This is in agreement with the original proposal by \mbox{\citet{Hoyle_Fowler}}. 

In the DD scenario, the focus of this work, Roche lobe overflow by the less massive star is expected to result in rapid accretion of material onto the more massive component. It is speculated that in the process, conditions amenable to runaway thermonuclear burning are achieved. There are multiple scenarios under which the ignition of the degenerate material is hypothesized to occur. Considering the time elapsed from the onset of mass transfer until ignition, one can differentiate between the models that ignite promptly \citep{pakmor+10, guillochon+10, dan+14, moll+2014}, at intermediate times when the bulk of accreted material still resides in a hot accretion disk around the primary \citep{mochkovitch+89}, or at late times when the remnant object undergoes compression in its core \citep{saio+85, vanKerkwijk+2010, raskin+2014}.

Recent evidence from sky surveys and population synthesis studies has reinforced interest in the DD channel. X-ray observations indicate that this scenario may be a preferred SN Ia production channel in early-type galaxies \citep{gilfanov+10}. Also, more recently, observations by the Nearby Supernova Factory suggest DD progenitors for five SN Ia events \citep{scalzo+2012}. These supernovae are believed to be super-Chandrasekhar in origin. Furthermore, observed distributions of SNe Ia time delays (the amount of time needed for a coeval stellar population to produce a Type Ia supernova) appear to favor the DD channel \citep{maoz+12, toonen+12}.

In this project, we study the white dwarf merger process at early times. Thus, we are limited to prompt ignition scenarios. As previous studies of such early merger stages indicated, in systems with components of differing mass, conditions amenable to ignition may occur in the hot boundary layer around the primary component \citep[see, e.g.,][]{dan+14}. The layer forms in the process of accretion of material  from the secondary, which is heated to high temperatures when the accretion stream passes through the accretion shock located above the boundary layer. Because angular momentum losses in the accreted material are negligible on timescales comparable to the orbital timescale, the accreted material remains rotationally supported at early times, as demonstrated by \citet{mochkovitch+89}. These authors considered a scenario in which accreted material forms a thick, turbulent disk. We note that turbulence is naturally generated during the accretion process. \citet{mochkovitch+89} speculated that ignition may occur at relatively low densities in the accreted material, either promptly (before accretion stops), or at later times when the disk material is heated to sufficiently high temperatures via turbulent dissipation. Their model accounted for the most significant components of the merger scenario and its evolution, but was unable to address in detail the ignition process itself. In particular, their model did not capture the effects of small-scale fluctuations in density and temperature that necessarily participate in the evolution. Such fluctuations are important due to a strong dependence of the burning timescale on temperature and, to a smaller degree, density \citep{dursi+06, malone+2014}. Also, the effects of stellar plasma degeneracy (or perhaps lack thereof) were not accounted for.

No evidence for successful explosions at early times was found in a number of carbon-oxygen white dwarf merger studies \citep[see, e.g.,][]{guerrero+04, yoon+07, loren-aguilar+09, dan+12}. However, as noted by \citet{pakmor+12aug}, hydrodynamic models utilized in these investigations lacked the resolution necessary to properly describe the system evolution on small scales--perhaps altering important simulation outcomes and conclusions. Additionally, given the constraints on the conditions required for triggering a self-sustained detonation \citep{dursi+06}, Smoothed Particle Hydrodynamics (SPH, \citet{lucy77, monaghan99, rosswog+2015}) models used in these studies would require a much greater number of particles in order to form and adequately resolve potential ignition points.

A number of more recent simulations reported conditions supportive of ignition in the case of high-mass carbon-oxygen systems \citep{pakmor+10, dan+14, pakmor+2011, pakmor+12mar, sato+2015, tanikawa+2015}. No naturally-occurring detonations were found in these studies. However, later studies combining SPH models with grid-based hydrocodes have produced self-sustained detonations \citep{kashyap+2015, moll+2014}. This is an encouraging development, although it is not clear how such hybridization of various hydrodynamic methods contributed to the simulation results. Ultimately, these findings should be scrutinized and confirmed using consistent models with well-known convergence properties and free of potential numerical artifacts due to data mapping between models using different mesh discretizations. 

Aside from challenges posed by numerics, one also faces the daunting task of selecting realistic physical models. For example, in systems with components of similar mass, a merger morphology is characterized by deep entraintment of each component's material into the other \citep[see, e.g.,][]{pakmor+10}. In this case, since the boundary layer per se does not form, ignition may occur in the central region of the forming merger object. Additionally, detonations may form under milder conditions in helium-rich mixtures \citep{holcomb+2013}. Indeed, a number of studies have indicated that ignitions are likely in helium-enriched boundary layers \citep{guillochon+10, pakmor+2013, shen+2014, tanikawa+2015, holcomb+2013}. The outcome of such ignition, however, does not necessarily guarantee a successful SN Ia explosion due to the low amount of energy released from helium burning \citep{Woosley+86}. Thus, various double-detonation scenarios are considered, in which the carbon-rich material of the primary component detonates. The two most popular scenarios here are as follows. In the first scenario, a detonation wave propagating through the helium-rich boundary layer directly ignites carbon-rich surface layers of the primary \citep{nomoto82b}. This is known as the edge-lit mechanism. Alternatively, provided that, prior to ignition, helium has been accreted in sufficient amounts in the equatorial region of the primary, it ignites at a point near the base of the accreted layer. The resulting detonation wave moves laterally through the helium layer and sends a converging shock into the interior of the primary, igniting carbon in its core \citep{livne90}. The double detonation scenarios were recently the subject of extensive studies in multidimensions \citep[see, e.g.,][]{fink+2007, guillochon+10,fink+10, sim+2012}. However, with the exception of work by \citet{guillochon+10}, these studies used a physical model that may not be suitable for white dwarf mergers.

In this work we present, for the first time, a set of white dwarf merger simulations solely performed using a grid-based hydrocode in three dimensions. We consider four binary systems with differing mass ratios and compositions, and total mass exceeding the Chandrasekhar mass. Our simulations start shortly before the onset of mass transfer and continue for several orbital periods. Since no model remapping is employed, our simulations are free of model perturbations possibly incurred in the remapping process \citep{pakmor+2013, moll+2014, kashyap+2015}. Because both stars are present on the mesh from the outset, we avoid the need for complex boundary conditions (e.g., to model the accretion stream \citep{guillochon+10}). Furthermore, we use adaptive mesh refinement to reduce numerical diffusion effects and probe evolution on smaller scales. In our analysis of simulation results, we pay close attention to conditions conducive to the detonation ignition and its subsequent evolution. In particular, we consider the mass of the ignition region, the importance of pre-processing via nuclear burning, and the role of degeneracy.

%% file: modelsMethods.tex
In this study, we performed a set of three-dimensional white dwarf merger simulations using \textsc{Proteus}, a fork of the \textsc{FLASH} code \citep{Fryxell+00}. \textsc{FLASH} is an adaptive mesh refinement multiphysics hydrocode. The adaptive nature of the mesh discretization allows for better control of solution accuracy and increased computational efficiency compared with simulations performed on a uniform mesh. Such savings are potentially very large in situations where regions of interest are small and few \citep{BC89}, as is the case in the present application. 

\subsection{Physics and Computational Models}

We solved the time-dependent reactive Euler equations for self-gravitating flow in Cartesian geometry in three spatial dimensions. The hydrodynamic evolution was calculated using the PPM solver \citep{colella+84}. We used a modified version of the \textsc{FLASH} multigrid elliptic solver to account for the effects of self-gravity, and the Helmholtz equation of state \citep{timmes+2000}. The nuclear burning was calculated using an alpha network with 19 isotopes\footnote{The corresponding matter composition was described using the following 19 nuclear species \citep{weaver+78}: \isotope[1]{H}, \isotope[3]{He}, \isotope[4]{He}, \isotope[12]{C}, \isotope[14]{N}, \isotope[16]{O}, \isotope[20]{Ne}, \isotope[24]{Mg}, \isotope[28]{Si}, \isotope[32]{S}, \isotope[36]{Ar}, \isotope[40]{Ca}, \isotope[44]{Ti}, \isotope[48]{Cr}, \isotope[52]{Fe}, \isotope[54]{Fe}, and \isotope[56]{Ni}, along with protons and neutrons.} The nuclear species were advected using the CMA method \citep{PM99}.   

To further minimize the effects of numerical diffusion due to rapid motion of fluid across the mesh in the rotating binary, in our simulations we used a co-rotating frame of reference. This approach also mitigates errors in angular momentum conservation. The process of the inspiral was not calculated self-consistently, i.e. we did not adjust orbital parameters based on the loss of angular momentum due to emission of gravitational waves. Rather, we forced both components to gradually inspiral by imposing a slightly sub-Keplerian orbital rotation. The rotation frequency, $\omega$, was a free parameter of our models. In addition, the velocity vectors of both components were slightly tilted in the orbital plane toward the system's center of mass by $0.005\times\pi$ \si{\radian}. Further details of our implementation of the  inspiral and discussion of the related effects are given in Section~\ref{s:orbitalParams}.
\subsubsection{Domain and Boundary Conditions}
The simulations used a 3D Cartesian geometry. To decrease the amount of time required for simulations, we assumed symmetry about the orbital plane. Therefore, only the $z>0$ half-space was considered. We allowed for free outflow through all boundaries, except across the orbital plane ($z=0$), at which we imposed a reflecting boundary condition. The simulation domain was a cube with a length of $\approx\SI{4.19e11}{\cm}$ in each direction. We found by conducting many preliminary simulations that such a large domain size was necessary in order to retain the mass on the mesh. Preserving the total mass in the computational domain is necessary for correctly capturing the system's dynamics (i.e., mass loss would result in a shallower gravitational potential, leading to artificial decompression of the stellar plasma). The domain was initially filled with pure helium at density \SI{e-3}{\g\per\cubic\cm} and temperature \SI{e6}{\K}, and suitable white dwarf models were created inside the domain using the procedure described in the following section.
\subsubsection{Initial Models}
In preparation for multidimensional studies, we created a database of spherically symmetric, isothermal white dwarf models with different masses and an initial temperature of \SI{e7}{\K}. For each multidimensional simulation, the structure of each component was initially interpolated onto the mesh with stellar centers placed according to the orbital distance such that the secondary was nearly filling its Roche lobe. The initial one-dimensional mass distribution was then adjusted iteratively using the Self-Consistent Field Method \citep[SCF,][]{hachisu+86} to account for rotational distortion of components and effects of self-gravity (see Sect.~\ref{s:selfConsistentField} below). From these initial conditions, we followed the evolution for several orbital periods. The simulations were typically terminated when either a successful detonation was launched and burning had essentially quenched, or ignition times in the simulations rose well above local dynamical timescales.
\subsection{Computational Methods}
In this section we briefly describe key modifications we made to the original \textsc{FLASH} code and included in \textsc{Proteus} to improve numerical accuracy and make merger simulations computationally feasible (the starting point for \textsc{Proteus} was version 4.0 of \textsc{FLASH}). These include changes to the burning modules, the multigrid and multipole solvers, and the implementation of a self-consistent field method to make initial conditions more consistent with the geometry of a close binary white dwarf system.

\subsubsection{LBNB: Load balancing for nuclear burning}
\label{s:loadBalance} 

Previous studies of phenomena involving thermonuclear transients \citep[see, e.g.,][]{plewa+07} indicated that intense localized burning may dominate computational time and induce substantial load imbalance, making multidimensional hydrodynamic studies computationally infeasible. In \citet{plewa+07}, regions of active thermonuclear burning were typically small and occupied only a few mesh cells. In consequence, the burning calculations were performed using only a small fraction of available MPI processes. This is because the computational workload in \textsc{Proteus} is distributed on a per-block basis. In the case  that there is a small cluster of only a few burning cells, it may so happen that the cluster will belong, to a single block. Therefore, burning calculations will be performed by only a single MPI process in this extreme case. Such situations will result in large load imbalances for massively parallel runs, in which only a few processes out of an allocated process set are involved in computations of nuclear burning. The already significant imbalance was further exaggerated by the high relative computational cost of burning as compared with other physics processes (several times more expensive than hydrodynamic advection).

The above-described deficiency of the commonly-used computational model of nuclear burning coupled to hydrodynamics motivated the development of a specialized load balancer for nuclear burning (LBNB). We begin by constructing a list of burning cells on each process. Along with this list, we also collect information about the cost of computing burning in these cells (measured as the CPU time). Then a global list of burning cells and their computational cost is constructed by gathering individual list data on the master process. In this way, we obtain information about the total computational cost of burning. The next task is to distribute this work evenly across available processes. That is, we create a number of sets of cells to be sent to processes participating in burning. The burning is then performed, and the results are distributed back to the original cell owners. The data exchanged in this process is rather small, requiring only information about the thermodynamic state and nuclear composition. We discuss the performance gains obtained thanks to load-balancing of nuclear burning in the appendix.
\subsubsection{Self-Gravity with Isolated Boundary Conditions}
\label{s:gravity}

In the original \textsc{FLASH} multigrid implementation of the Poisson equation solver for self-gravity, the solution residuals were computed using the current defect and correction values. (For details of the multigrid method, please consult \citet{briggs+2000}; for implementation details, see \citet{ricker2008}.) This approach results in a slow numerical drift of the solution, most likely due to details of the treatment of boundary conditions. We found that this problem can be remedied by computing residuals directly from the source and solution. This approach leads to slightly different but more accurate results.

The convergence of the multigrid solver can also be improved by providing a more accurate initial guess for the solution by extrapolating data in time using solutions from previous timesteps. For example, an initial guess computed using linear extrapolation usually saves one or two V-cycles, which reduces the time to solution by about 30 percent.

We also found that under certain conditions (many levels of refinement, complex refinement topology, frequent mesh refinement/derefinement), mesh resolution may change by more than one refinement level across block boundaries in the diagonal direction. This rarely-occuring error of the \textsc{FLASH} \textsc{PARAMESH}-based \citep{macneice+2000} meshing package resulted in strong variations of the solution defects between iterations. In consequence, the convergence was either substantially slower, or the solution could not be obtained. As a workaround for this problem, we modified the refinement scheme in such a way that mesh resolution could change only by one level between diagonally-adjacent blocks.

Furthermore, we lowered the memory footprint of the multipole solver by reorganizing memory access. Also, we provided for better handling of floating-point overflows and underflows in the calculation of Legendre coefficients in the multipole solver. These optimizations also slightly decreased the computational cost of obtaining isolated boundary conditions for the multigrid solver. More importantly, we optimized a subset of the original \textsc{FLASH} \textsc{PARAMESH} MPI communication library modules, providing a speedup by a factor of 3 under conditions typical to our simulations.
\subsubsection{Self-Consistent Field Method}
\label{s:selfConsistentField}

\citet{dan+11} have demonstrated that the dynamical timescale of the merger is severely underestimated in models initialized with ``approximate'' initial conditions in which the model stars are spherically symmetric and placed at a distance corresponding to a binary system with the secondary nearly overflowing its Roche lobe. These authors advocated using ``accurate'' initial conditions in which the components' geometries accounts for self-gravity, and tidal and centrifugal effects. To this end, they used a scheme that iteratively drives the particle distribution in their SPH simulations to a dynamical equilibrium configuration. Although the same approach could be used in the case of the grid-based simulations presented here, a method specifically designed to construct equilibrium models of rotating objects was introduced by \citet{hachisu+86}. Hachisu's Self-Consistent Field method (SCF) uses a corotating frame of reference and starts with spherically-symmetric white dwarf models interpolated onto a mesh at a distance consistent with the orbital period. Next, the density field is iteratively relaxed toward an equilibrium state. Each iteration of SCF involves first computing the gravitational potential associated with the current density field (due to the stellar components). In our \textsc{Proteus} implementation, the self-gravitating potential is calculated using the modified multigrid solver described in the previous section. The SCF solver then iteratively adjusts the mass (density) distribution within equipotential surfaces corresponding to the masses of the components. In the last step of the iteration, the self-gravitating potential is updated based on the modified density field. The solver usually converges in about 20 iterations before the correction of the potential on the stars' surfaces decreases below a specified relative error. The mesh is automatically adapted throughout this process (see the following section).
\subsubsection{Adaptive Mesh Refinement}
One of the key features of \textsc{Proteus} is its adaptive mesh refinement (AMR) capability. AMR allows for better control of the solution error and offers potentially large savings in computer time and memory usage, especially for problems with localized flow discontinuities \citep{BC89} or source terms \citep{kapila+2002}. For example, in the current application, it is important to resolve steep gradients associated with surface layers of white dwarfs, the accretion shock, the hot boundary layer, and regions of intense nuclear burning. In the course of the simulation, these structures can be identified using a set of specialized refinement criteria and flagged for refinement.

We used the following set of refinement criteria. A block was marked for refinement if the density and temperature in any cell in that block exceeded \SI{1.5e5}{\g\per\cubic\cm} and \SI{4e9}{\kelvin}. In addition, we refined blocks with density contrasts in excess of 0.5, provided the density and pressure in any of their cells exceeded \SI{e3}{\g\per\cubic\cm} and \SI{3e20}{\dyn\per\square\cm}, respectively. Also, refinement was triggered if the minimum ignition time, defined as the time required for a parcel of stellar plasma to initiate runaway burning \citep{dursi+06}, in any cell inside a block was less than \SI{e4}{\s}. A minimum mesh resolution of \SI{256}{\km} was enforced for all blocks whose density was greater than \SI{e3}{\g\per\cubic\cm}. Finally, no block located at distances greater than \SI{3e9}{\cm} from the system's barycenter could have a resolution better than \SI{512}{\km}. 

\subsection{Model Parameter Space} \label{s:parameterSpace}
In this work, we studied a select number of binary white dwarf systems differing in mass ratio and progenitor composition. We chose mass ratios to allow for comparison with existing results. Specifically, we considered carbon/oxygen (CO) binary systems with component masses of $0.6 + \SI{0.9}{\solarmass}$ \citep[model CO0609; see, e.g.,][]{guerrero+04} and $0.895 + \SI{0.905}{\solarmass}$ \citep[model CO0909; see, e.g.,][]{pakmor+10}. In this study, CO white dwarfs are made of equal mass fractions of carbon and oxygen. We also considered one high-mass CO system with component masses of $0.8 + \SI{1.2}{\solarmass}$ (model CO0812). We investigated the effects of composition in an additional $0.6 + \SI{0.9}{\solarmass}$ system with a pure helium secondary (model He0609). Note that the total mass of the systems used in our study always exceeded the Chandrasekhar mass, potentially enabling long-term evolution studies of the remnant \citep[see, e.g.,][]{raskin+12}. The initial binary separation $d_0$, the corresponding initial orbital period $P_0$ of the systems, and the adopted inspiral rates are given in Table~\ref{t:parameterSpace}. 
\ctable[
cap = orbital parameters,
caption = {Masses, composition, and initial orbital parameters of binary merger models.\tmark} ,
label = t:parameterSpace,
star,
]{cccccc}
{
\tnote{All models were run at an effective resolution of \SI{64}{\km}.}
\tnote[b]{$f_K=\omega /\omega_K$, where $\omega_{K}$ is the Keplerian orbital frequency at the initial time.}
}
{\FL
Model       & Progenitor                                & Progenitor    & $d_0$          & $P_0$ & $f_K$\tmark[b]    \NN
designation & masses $\left(\mathrm{M_{\odot}}\right)$  & composition   & (\SI{e9}{\cm}) & (s)   &          \ML
He0609      & \multirow{2}{*}{$0.6+0.9$}                & He/CO         & 2.9            & 69.4  & 0.999    \NN
CO0609      &                                           & CO/CO         & 2.9            & 69.4  & 0.999    \NN
CO0812      & $0.8+1.2$                                 & CO/CO         & 2.5            & 48.1  & 0.999    \NN
CO0909      & $0.9+0.9$                                 & CO/CO         & 1.8            & 30.1  & 0.99     \LL  
}
The justifications for specific choices of the orbital parameters, including the inspiral rates, are discussed in Sec.~\ref{s:orbitalParams}. In passing, we note that the assumed composition of our white dwarf models is simplified. For example, a 0.6 solar mass white dwarf is expected to be chiefly composed of carbon and oxygen, with only a very small helium envelope. In addition, our exploratory stellar evolution calculations performed with the \textsc{Mesa} code \citep{paxton+2015} point to substantial amounts of neon present in the cores of massive white dwarfs, with the carbon abundance reduced by a factor of about 3 for a 1.2 solar mass white dwarf. Although the composition is not expected to play a decisive role in the dynamical evolution of mergers, it will influence the nuclear burning. Compositionally-realistic merger models would most likely be characterized by lower levels of heating due to nuclear burning and longer ignition times. This problem is beyond the scope of the present work. 

As mentioned above, the nearly equal-mass progenitor scenario was motivated by the work of \citet{pakmor+10}. Although these authors mentioned that the components of their model system differed slightly in mass, they did not quantify that difference. Subsequently, we were unable to specify model parameters to exactly match \citet{pakmor+10}, and our choice of $0.895 + \SI{0.905}{\solarmass}$ components is rather arbitrary. Perhaps more importantly, we performed simulations using an AMR grid-based code, while \citet{pakmor+10} used an SPH method. These methods are known to have very different convergence properties for discontinuous flows \citep[see, e.g.,][and references therein]{hubber+2013}. In passing, we note that we performed several trial simulations with components of exactly equal mass, but found a negligible difference in results from the non-equal-mass model.

For every binary system model, we executed a series of simulations, gradually increasing the effective mesh resolution, $\Delta x_{\text{eff}}$, starting at \SI{512}{\km}. We found that large-scale flow features (accretion stream, accretion shock, boundary layer) first became resolved when the effective resolution reached \SI{128}{\km}. Next we obtained complete simulations at an effective resolution of \SI{64}{\km}, which allowed resolving mixing instabilities in the boundary layer for the first time in our simulations. Subsequently, we adopted this resolution as the starting model resolution in our studies. As the merger process advanced, we periodically adapted the mesh primarily to better resolve material heated by tidal effects, and by the accretion shock. Substantial resolution is also required at later times to resolve fluid flow instabilities (such as Kelvin-Helmholtz), material mixing, and fluctuations in density and temperature that may play an important role in burning. 

%% file: results.tex
We present here the results of the simulations for each of the scenarios listed in Table~\ref{t:parameterSpace}. This series of simulations was obtained with a nominal effective mesh resolution of \SI{64}{\km}, and using an alpha network with 19 isotopes. We discuss the dependence of the results on the choice of nuclear network in Section~\ref{s:sensitivityNuclearNetwork}. We begin our presentation with the CO0609 model, which displays the most generic morphological features of unequal-mass white dwarf mergers. To make our results more accessible to the reader, a set of digital movies illustrating the evolution of our merger models is available for download at \url{http://people.sc.fsu.edu/~tplewa/WDmergers/index.html}.  
\subsection{C/O $0.6 + \SI{0.9}{\solarmass}$ Merger \label{CO0609results}}
\begin{figure*}
\adjustThreePanels\ignorespaces
\begin{center}
    \begin{tabular}{ccc}
        %
        \includegraphics{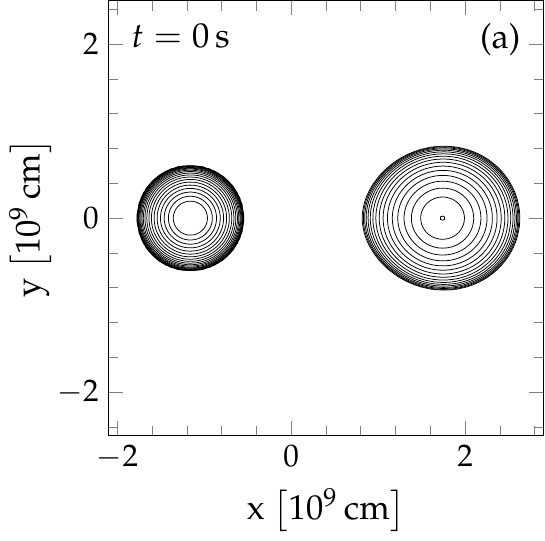}
        &
        \includegraphics{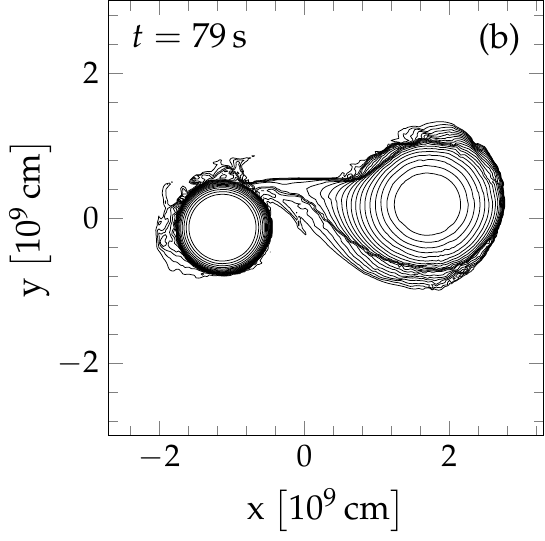}
        &
        \includegraphics{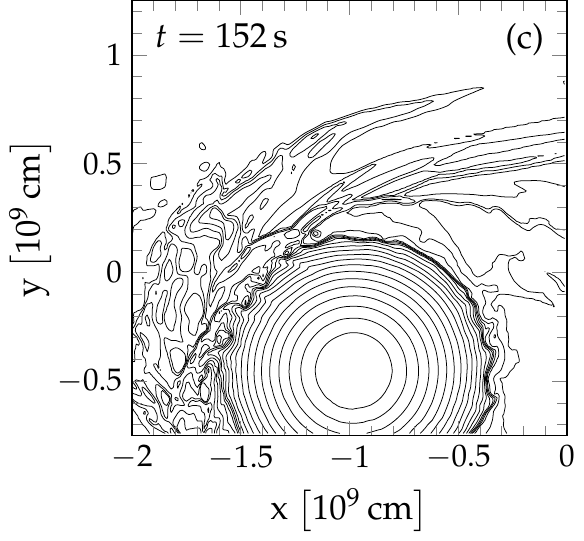}
        \\
        %
        \includegraphics{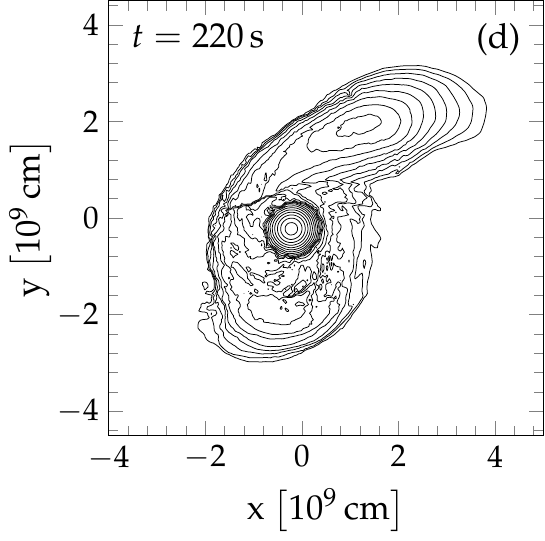}
        &
        \includegraphics{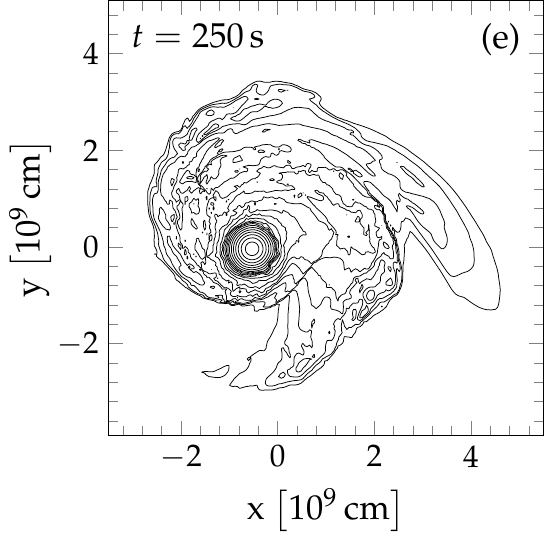}
        &
        \includegraphics{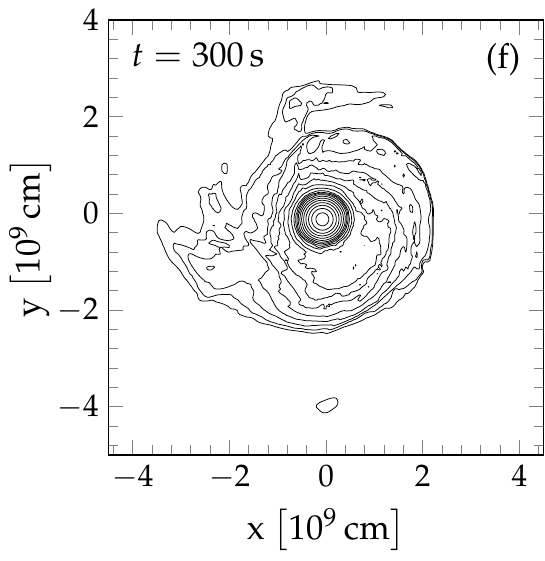}          
    \end{tabular}
    \caption
    {
        The CO0609 merger model. The density distribution in the model orbital plane is shown with 20 logarithmically-spaced contour lines between \num{1e4} and \SI{1e7}{\g\per\cubic\cm}, except for panel (b), in which the contours span the interval between \num{1e2} and \SI{1e6}{\g\per\cubic\cm}. Note that the region shown changes between the panels. See text for discussion. \label{f:mergerMorph_0609}
    }
\end{center}
\end{figure*}

The morphology of the CO0609 model is shown with density contours in a series of panels in Figure~\ref{f:mergerMorph_0609}.
%
%
%
The initial conditions in this model are shown in Figure~\ref{f:mergerMorph_0609}(a). At this time, the secondary shows modest tidal distortion, while the primary remains essentially spherically symmetric. After about one orbital period, the mass transfer is significantly advanced (Figure~\ref{f:mergerMorph_0609}(b)), with the accretion stream striking directly on the surface of the primary. The bulk of accreted material is confined in a stream several hundred kilometers wide. The accreted material splashes on the surface and moves around the primary, initiating the formation of a boundary layer. After around 2.5 orbits (Figure~\ref{f:mergerMorph_0609}(c)), the boundary layer is well-developed, with the outer layers of the primary being deformed due to Kelvin-Helmholtz instabilities (KHI) with a characteristic wavelength of about \SI{1000}{\km}. Given our mesh resolution, these structures appear well-resolved. The overall morphology after around three orbital periods is shown in Figure~\ref{f:mergerMorph_0609}(d). At this time, the secondary is completely disrupted and presents a tidally distorted blob of material. The matter accreted from the secondary now passes through the accretion shock, visible as closely-spaced density contours extending nearly horizontally around $y\approx\SI{4e8}{\cm}$ and from $x\approx\num{-1.6e9}$ to $x\approx{\SI{4e8}{\cm}}$. Note that the boundary layer at this time is approximately elliptical in shape as the material accreted from the secondary carries angular momentum that remains to be redistributed during the process of the disk formation. This process is illustrated in the last two panels in Figure~\ref{f:mergerMorph_0609}. During this time, the secondary is completely absorbed, and the accreted material settles down in its Keplerian motion around the primary. No detonation was triggered in this model.
\subsection{$0.6$ He + $\SI{0.9}{\solarmass}$ C/O Merger \label{He0609results}}
\begin{figure*}
\adjustThreePanels\ignorespaces
\begin{center}
    \begin{tabular}{ccc}
        %
        \includegraphics{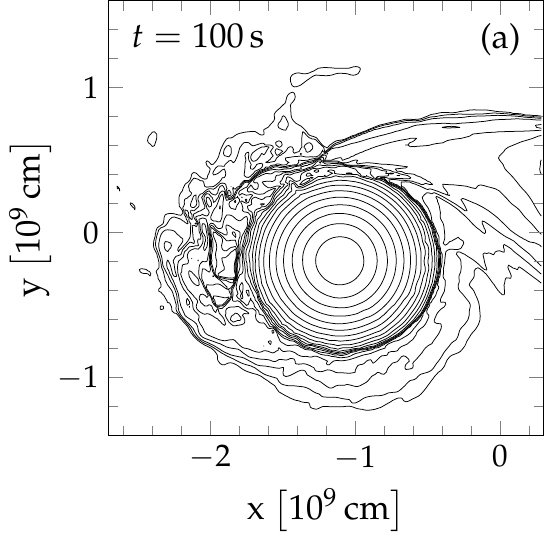}
        &
        \includegraphics{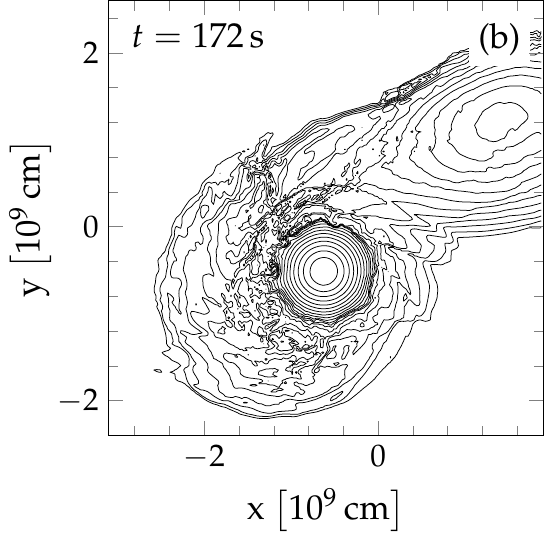}
        &
        \includegraphics{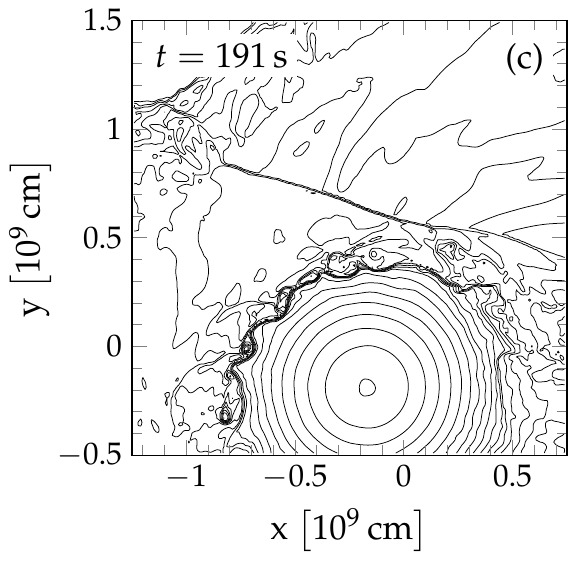}             
    \end{tabular}
    \caption
    {
        The violent phase of the He0609 merger model. The density distribution in the model orbital plane is shown with 20 logarithmically-spaced contour lines between \num{1e4} and \SI{1e7}{\g\per\cubic\cm}, except for panel (a), in which contours span the interval between \num{1e3} and \SI{1e7}{\g\per\cubic\cm}. Note that the region shown changes between the panels. See text for discussion. \label{f:mergerMorph_0609he}
    }
\end{center}
\end{figure*}

\begin{figure*}
\adjustThreePanels\ignorespaces
\begin{center}
    \begin{tabular}{ccc}  
        %
        \includegraphics{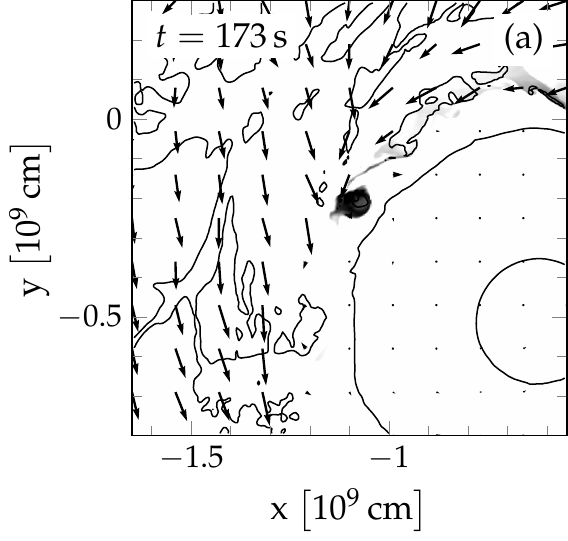}
        &
        \includegraphics{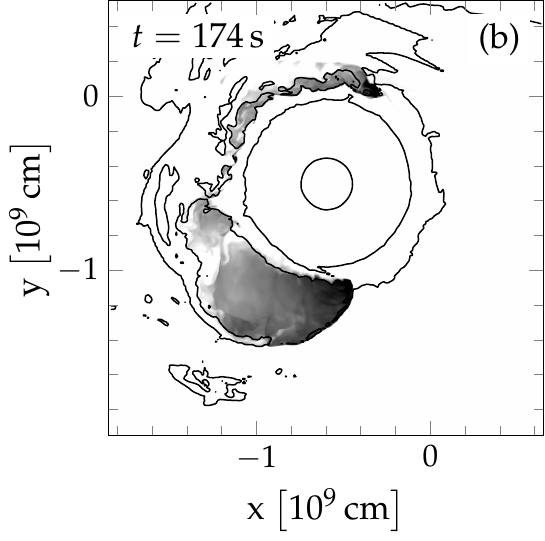}
        &
        \includegraphics{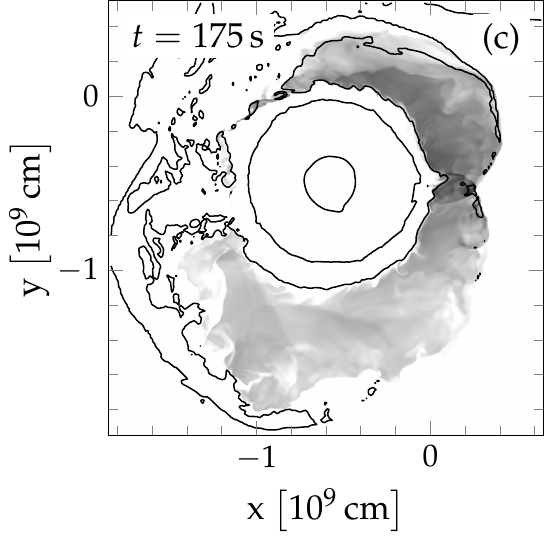}            
    \end{tabular}
%
    \includegraphics{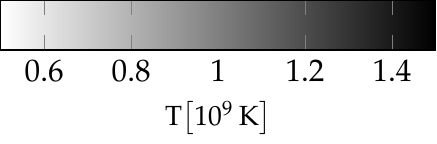}
    \caption
    {
        The evolution of the helium detonation in the He0609 merger model. The temperature distribution in the model orbital plane is shown with a pseudocolor map in linear scale for temperatures above \SI{0.5e9}{\K}. The upper temperature limit in the plot, \SI{1.5e9}{\K}, was chosen such that the temperature variations are well-visible at all times. (The maximum recorded temperatures in this model approached \SI{3e9}{\K}.) Contours represent densities of $10^5$, $10^6$, and \SI{e7}{\g\per\cubic\cm}. The velocity field is shown in panel (a), where vectors with a maximum length correspond to $\lvert v \rvert=\SI{5e8}{\cm\per\s}$. Note that the region shown changes between the panels. See text for details. \label{f:heliumDet}
    }
\end{center}
\end{figure*}

The overall evolution in the case of the helium secondary for the $0.6 + \SI{0.9}{\solarmass}$ case is qualitatively similar to the C/O case discussed above. Thus, here we restrict our description of the evolution to intermediate times depicted in the series of panels in Figure~\ref{f:mergerMorph_0609he}.
%
The density structure around the primary at $t=\SI{100}{\s}$ is shown in Figure~\ref{f:mergerMorph_0609he}(a). At this time, we noticed the early development of Kelvin-Helmholtz instabilities at the surface of the primary. Similarly to the C/O binary case, this instability is triggered by a shear between the secondary material accreted in the boundary layer and the surface layers of the primary. The resulting structure is illustrated in Figure~\ref{f:mergerMorph_0609he}(b). By this time, the boundary layer, fed through the accretion shock located a few hundred kilometers above the surface of the primary, has grown significantly in mass and thickness. The structure of the layer is complex, with the flow perturbed by both the KHI and the interaction of the perturbed boundary layer material with the accretion stream. The flow perturbations appear strongest when the shocked accretion stream collides with the boundary layer. The accretion shock can be seen as the closely-spaced density contour lines near $(x,y)=(\num{-1e9},\num{2e8})\, \si{\cm}$ in Figure~\ref{f:mergerMorph_0609he}(b) (rapid changes in the velocity field associated with the shock can be seen in the same region in Figure~\ref{f:heliumDet}(a)). At this time, a small, helium-rich region located downstream from the accretion shock near $(x,y)=(\num{-1.1e9},\num{-0.2e9})\, \si{\cm}$ detonates, and can be seen as an area of temperatures in excess of \SI{1.5e9}{\K} in Figure~\ref{f:heliumDet}(a).
%
%
We defer discussion of the details of the detonation process to Section~\ref{s:detonationsHeliumBL} below. 

The detonation wave moves away from the detonation point, propagating through the boundary layer in a highly non-steady fashion as it encounters the perturbed boundary-layer material. This is illustrated in Figure~\ref{f:heliumDet}(b), in which the hot material burned by the clockwise-moving detonation is shown to occupy a relatively narrow and meandering channel. The lack of lateral expansion of this burned material can be explained by the confinement provided by the incoming shocked accretion stream. As soon as the detonation exits this region, it is able to expand laterally, as can be seen in Figure~\ref{f:heliumDet}(c). In contrast, the material burned by the counterclockwise-moving detonation is not subjected to such confinement, and can expand uninhibited as the detonation moves through the boundary layer (seen in the lower section of Figure~\ref{f:heliumDet}(b)). Eventually, the counterclockwise-moving detonation encounters the clockwise-moving front. We note that the leading edges of both detonation fronts move through the base of the boundary layer, where the conditions for explosive helium burning are the most favorable (i.e., the ignition time for helium-rich material is the shortest, thanks to its relatively high density). We found no evidence for explosive carbon burning in this model. At later times, the evolution proceeds in a fashion very similar to that in the C/O case discussed above (cf.\ Figure~\ref{f:mergerMorph_0609}(c)) with pronounced KHI structures developing most vigorously at the portion of the primary's surface exposed to the incoming shocked stream material (Figure~\ref{f:mergerMorph_0609he}(c)).
\subsection{$0.8 + \SI{1.2}{\solarmass}$ Merger} \label{s:results0812}
\begin{figure*}
\adjustThreePanels\ignorespaces
\begin{center}
    \begin{tabular}{ccc}
        %
        \includegraphics{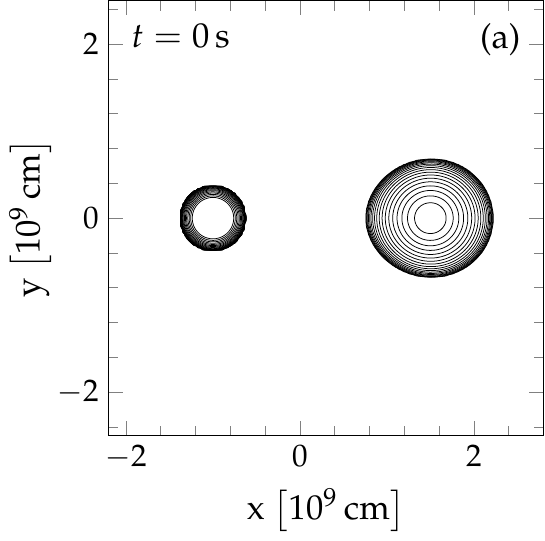}
        &
        \includegraphics{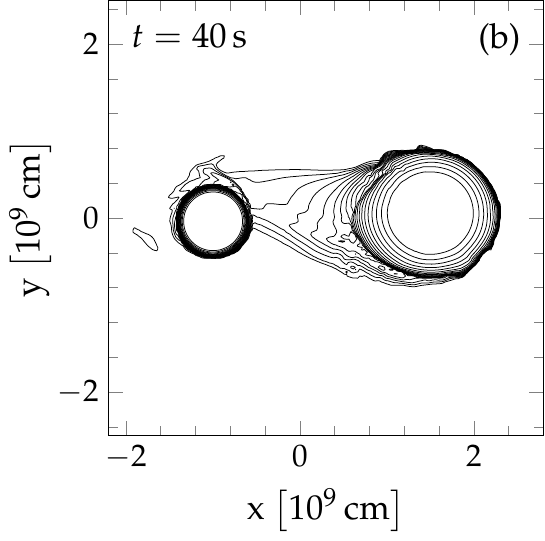}
        &
        \includegraphics{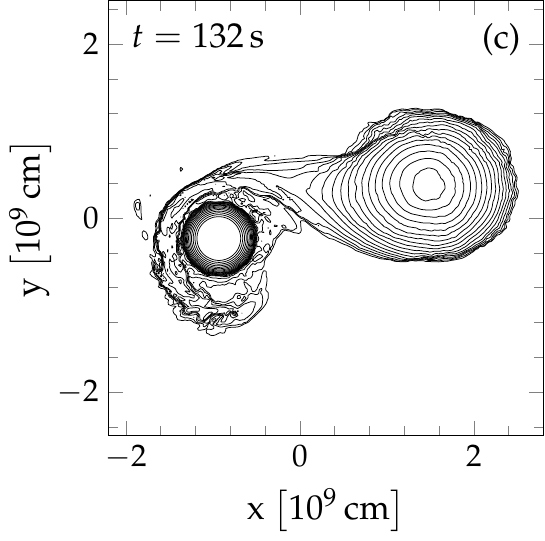}  
        \\
        \includegraphics{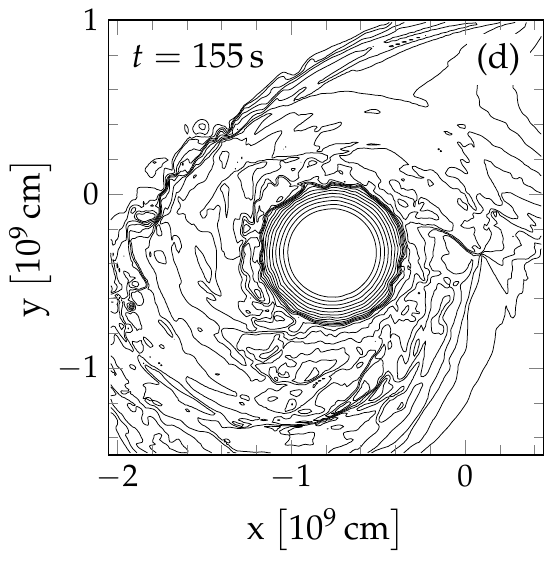}
        &
        \includegraphics{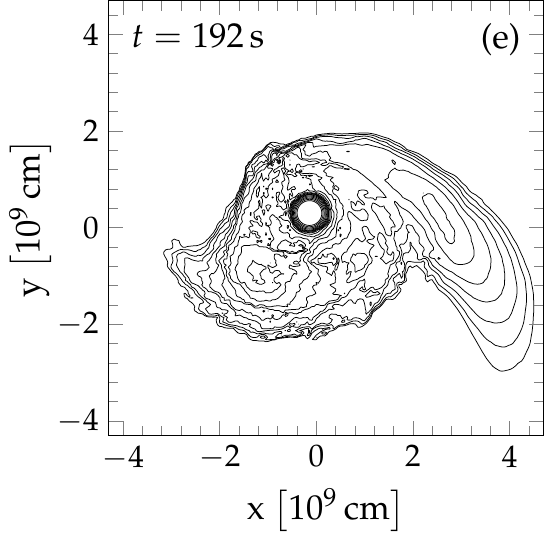}
        &
        \includegraphics{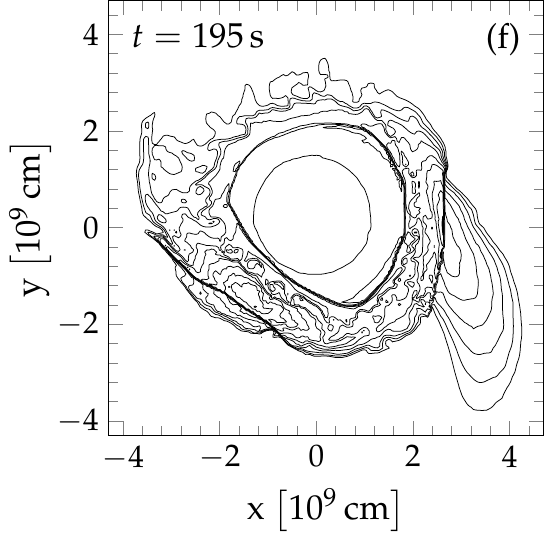}           
    \end{tabular}
    \caption
    {
        The CO0812 merger model. The density distribution in the model orbital plane is shown with 20 logarithmically-spaced contour lines between \num{1e4} and \SI{1e7}{\g\per\cubic\cm}, except for panel (b), in which the contours span the interval between \num{1e1} and \SI{1e6}{\g\per\cubic\cm}. Note that the region shown changes between the panels. Panel (e) shows the model at the time of detonation ignition in the core of the primary. The structure of the exploding model is depicted in panel (f). See text for discussion. \label{f:mergerMorph_0812}
    }
\end{center}
\end{figure*}

The second C/O binary we considered, CO0812, was comprised of somewhat heavier components, in which case one may expect more violent evolution due to deeper potential wells. For the adopted initial conditions, the secondary component appears more deformed (Figure~\ref{f:mergerMorph_0812}(a))
%
than in the CO0609 case (cf.\ Figure~\ref{f:mergerMorph_0609}(a)). Here, the first noticeable effects of mass transfer can be seen at around $t=\SI{40}{\s}$ (Figure~\ref{f:mergerMorph_0812}(b)). At this time, a thin layer of shocked accreted material can be seen downstream from the region impacted by the accretion stream. As in the CO0609 case, a boundary layer forms from the accreted material (Figure~\ref{f:mergerMorph_0812}(c)), but it takes relatively longer (three orbital periods) for the Kelvin-Helmholtz instability to develop at the surface of the primary. Figure~\ref{f:mergerMorph_0812}(d) shows the density distribution at $t=\SI{155}{\s}$, with the boundary layer exhibiting a strongly-perturbed morphology, including several shocklets; the characteristic wavelength of KHI is $\approx\SI{1500}{\km}$. Although the boundary layer displays the highest temperatures found in this model at this time, and we observed several pockets of intense burning, no detonation was triggered in the boundary layer. Instead, it is the dense core region of the primary where a combination of compressional heating due to accretion and degeneracy effects conspire to initiate a detonation. The morphology of the model around the time of detonation, $t=\SI{192}{\s}$, is depicted in Figure~\ref{f:mergerMorph_0812}(e). At this time, the core of the primary shows relatively little variation in the overall density distribution, with typical densities $\approx\SI{6e7}{\g\per\cubic\cm}$, while the secondary is strongly disrupted and forms a tidal tail. We discuss details of the ignition process and the later evolution in Section~\ref{s:coreCarbonIgnition} below.

We followed the evolution of the explosion for about 3.5 seconds after the detonation was triggered, when the detonation entered the low-density regions and nuclear burning quenched. The density structure of the central region in our model is shown in Figure~\ref{f:mergerMorph_0812}(f). A number of distinct features can be noted. The innermost portion of the merger remnant is characterized by a smoothly-varying density distribution, which gradually decreases from the center. This is the shocked material of the primary. It is bounded by a reverse shock whose surface at this time is roughly triangular in shape (in the model orbital plane), with one of the triangle's apexes located near $(x,y)=(\num{8e8},\num{-1.6e9})\, \si{\cm}$. The reverse shock formed after the supernova shock left the bulk of the primary and entered the boundary layer. The change in the density gradient from steep (inside the bulk of the primary) to shallow (in the boundary layer) caused the supernova shock to decelerate. In the process, the faster-moving shocked material of the primary began to collide with the slower-moving shocked boundary layer material. This collision process eventually produced a reverse shock. (In passing, we note that this mechanism of reverse shock formation also operates in core-collapse supernovae, where the supernova shock moves through the progenitor's envelope and encounters varying density gradients \citep[see, e.g., Section 6.1 in][]{kifonidis+03}.) Therefore, the convoluted density field visible outside the reverse shock front represents the remnants of the boundary layer. Finally, the supernova shock can be seen expanding past the boundary layer material, with the exception of the dense part of the boundary layer near $(x,y)=(\num{-2e9},\num{-1.6e9})\, \si{\cm}$ and the tidal tail of the secondary near $(x,y)=(\num{2.6e9},\num{-4e8})\, \si{\cm}$. While the shock has moved into the low-density region in the lower-right portion of the image only recently, it has already broken out of the lower-density section of the boundary layer in the top portion of the merger remnant. Thus, at the final time in our model, the supernova shock is highly asymmetric, and the expansion is far from homologous. 

\begin{figure*}
\adjustThreePanels\ignorespaces
\begin{center}
    \begin{tabular}{ccc}
        %
        \includegraphics{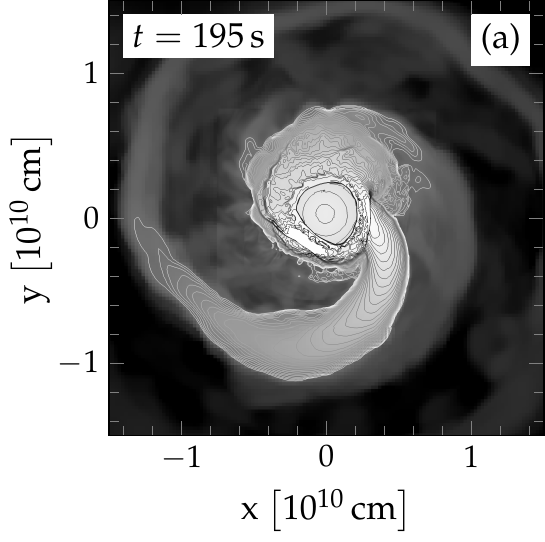}
        &
        \includegraphics{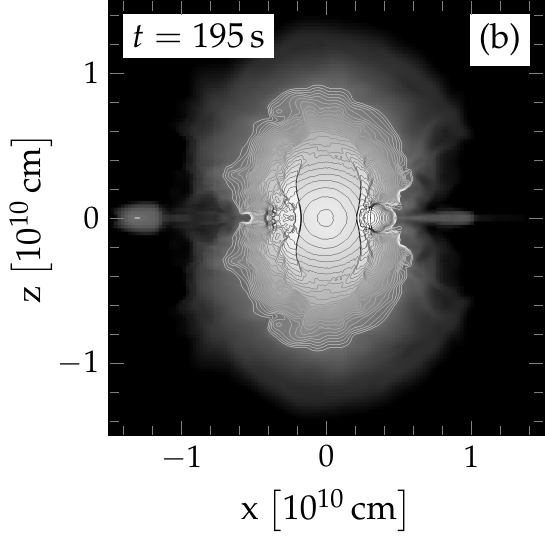}
        &
        \includegraphics{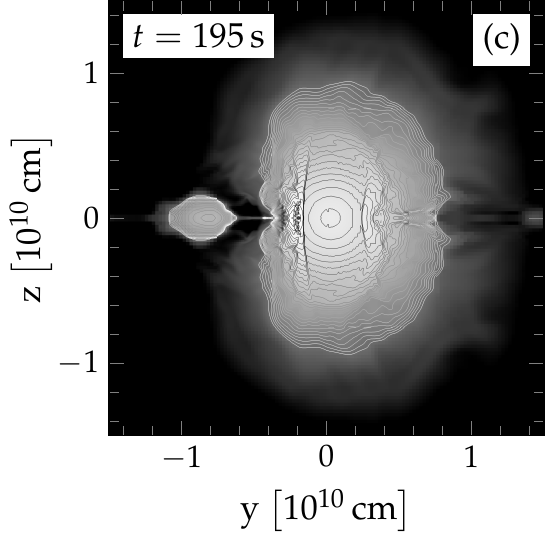}           
    \end{tabular}
%
    \includegraphics{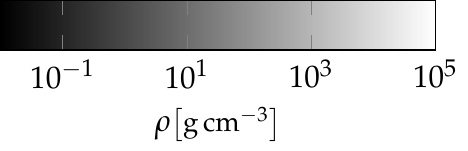}
    \caption
    {
        The CO0812 merger model. The density distribution is shown with a pseudocolor map and with 30 logarithmically-spaced contour lines between \num{1e1} and \SI{1e5}{\g\per\cubic\cm} in (a) orbital plane, (b) x-z plane, and (c) y-z plane at $t=\SI{195}{\s}$. See text for discussion. \label{f:mergerMorph_0812_vertical}
    }
\end{center}
\end{figure*}

It is important to note that the ejecta is not only aspherical in the model orbital plane, but is also elongated in the direction perpendicular to the model orbital plane. This aspect of our model is illustrated in Figure~\ref{f:mergerMorph_0812_vertical}.
%
%
Although the central part of the ejecta appears spherically-symmetric, the flow is restricted in the model orbital plane due to the presence of a substantial amount of accreted material (see panels (b) and (c) in Figure~\ref{f:mergerMorph_0812_vertical}). Furthermore, the expansion will continue to be restricted near the model orbital plane for the next several hundred seconds due to the presence of material from the disrupted secondary at high radii in the form of tidal tails (see panels (a) and (c) in Figure~\ref{f:mergerMorph_0812_vertical}). The estimated aspect ratio of the ejecta for density contours of $\approx \SI{100}{\g\per\cubic\cm}$ is around 2, although this may not reflect the asymmetry of the homologously-expanding ejecta, as relatively strong non-radial pressure gradients are still present at this time.
\subsection{$0.9 + \SI{0.9}{\solarmass}$ Merger}
As our final model, we considered a system composed of C/O white dwarfs with masses of $\SI{0.895}{\solarmass}$ and $\SI{0.905}{\solarmass}$. Our choice of the masses was motivated by the work of Pakmor and collaborators, who considered similar systems with only slightly less massive components (masses of $\approx\SI{0.89}{\solarmass}$ in \citet{pakmor+10}, and masses of $\approx\SI{0.9}{\solarmass}$ in \citet{pakmor+2011}). Perhaps more importantly, however, their systems appeared to be very tight, with components slightly more than 10 percent closer than the tightest system we considered. As we will show in Sec.~\ref{s:orbitalParams}, the initial orbital configuration does play a very important role in the process of merger, especially in terms of the temperature evolution of the accreted material. Here we limit our presentation to a description of the evolution of our CO0909 model.

\begin{figure*}
\adjustThreePanels\ignorespaces
\begin{center}
    \begin{tabular}{ccc}
        %
        \includegraphics{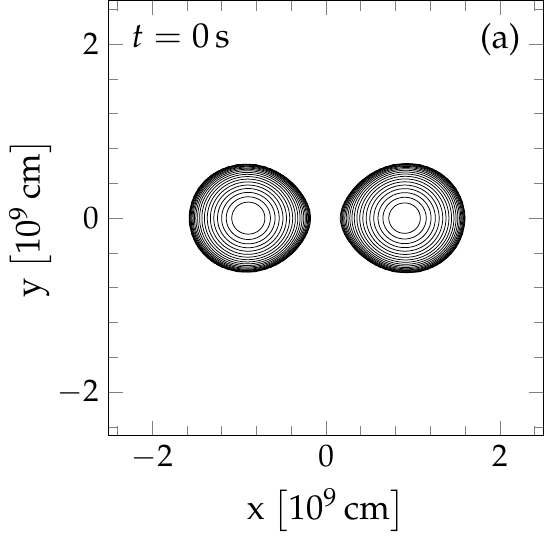}
        &
        \includegraphics{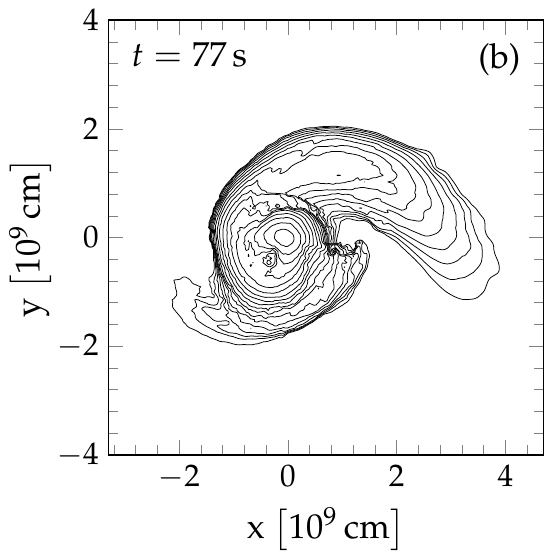}
        &
        \includegraphics{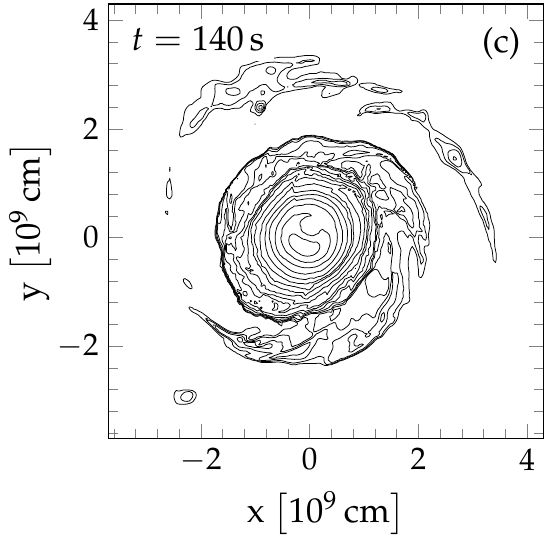}             
    \end{tabular}
    \caption
    {
        The CO0909 merger model. The density distribution in the model orbital plane is shown with 20 logarithmically-spaced contour lines between \num{1e4} and \SI{1e7}{\g\per\cubic\cm}. Note that the region shown changes between the panels. See text for discussion. \label{f:mergerMorph_0909}
    }
\end{center}
\end{figure*}

For an orbital separation of $d_0=\SI{1.8e9}{\cm}$, the white dwarfs are well-separated but visibly tidally-disorted at the initial time, as shown in Figure~\ref{f:mergerMorph_0909}(a).
%
%
In this case, the mass transfer begins after almost one orbital period, but the following accretion does not lead to the formation of the distinct boundary layer observed in model systems with components substantially differing in mass. In the present case, the accretion stream takes the form of a dense tongue of material from the secondary that penetrates the outer layers of the primary. Thus, the accreted material does not completely encircle the primary, as was the case in models discussed previously, but instead is quickly halted by the dense, intermediate-mass layers of the primary. Because the accreted material carries a substantial amount of angular momentum, the contact surface separating the primary and the secondary material is subject to extremely strong, persistent shear. At higher radii, the stream remains completely engulfed in the primary material. The leading edge of the accretion stream forms a vortex that is continually fed by the accreted material (the vortex core can be seen near $(x,y)=(\num{-4e8},\num{-4e8})\, \si{\cm}$ in Figure~\ref{f:mergerMorph_0909}(b)). At the same time, the core of the primary is deformed into a kidney-like shape. Due to the difference of the specific angular momentum between the vortex material and the primary's core, both structures rotate at differing rates. While the main effects due to the presence of the vortex are generation of turbulence and mixing, the primary's rotating core modulates the background density in the vicinity of the vortex. We stopped our simulation at $t=\SI{140}{\s}$, shown in Figure~\ref{f:mergerMorph_0909}(c), when the merger appeared complete, with only a tenuous tidal tail remaining from the secondary. By this time, the composite core of the merger consists of the vortex filled with material from the secondary, which is almost completely engulfed by the core material of the primary. We found no evidence for explosive burning in this model.

%% file: discussion.tex
In this section we discuss critical aspects of our set of merger simulations. This includes a detailed description of observed detonations, the structure of our models at final times, and the results of a systematic study of the sensitivity of simulation outcomes to model parameters.
\subsection{Conditions for ignition}
In assessing the likelihood of ignition in the context of the present work, a useful metric is the ignition time of stellar plasma for conditions considered here. Arising from the strong temperature dependence of thermonuclear burning at high densities \citep[see, for example,][]{woosley+04}, the ignition time represents the amount of time required for a parcel of stellar plasma to initiate runaway burning \citep{dursi+06}. In order to prevent substantial cooling of the burning region due to hydrodynamic expansion, the burning process must occur on a timescale shorter than the local dynamical timescale, $\tau_\mathrm{dyn}=446\rho^{-1/2}$ \citep{Fowler+64,Arnett96}. In the formula, $\rho$ is the density in cgs units, and the dynamical timescale is given in seconds. After considering additional factors, such as dynamics of the hydrodynamic background, one can estimate the likelihood of a detonation forming. 

Because typical densities found in our models in regions where the burning is active are lower than the lowest considered by \mbox{\citet{dursi+06}}, $\rho=\SI{1e7}{\g\per\cubic\cm}$, we extended their study to densities down to \SI{1e5}{\g\per\cubic\cm}. To this end, we performed a series of single-zone simulations following the procedure used by \mbox{\citet{dursi+06}}. In the process, we found that their Equation (1)  underestimates the ignition times around $(\rho,T)=(\num{1e5},\num{1e9})$ by a factor of about 4, consistent with their overall claimed accuracy. In the case of helium burning, we used the results of \citet[][Equation (A3)]{khokhlov+86}. 

An important factor controlling the explosiveness of burning stellar plasma is its degree of degeneracy \mbox{\citep[see, e.g., ][]{Hoyle_Fowler}}. To better understand the role of degeneracy in our simulations, we calculated the contribution of the thermal pressure component to the total plasma pressure for a mixture composed of equal mass fractions of carbon and oxygen. We used the Helmholtz equation of state and performed a series of single-zone calculations, varying the plasma temperature from $T=\SI{1e8}{\K}$ up to $T=\SI{1e10}{\K}$, and the density from $\rho=\SI{1e5}{\gram\per\cubic\cm}$ to $\rho=\SI{2e8}{\gram\per\cubic\cm}$. Then the ratio of pressure at a given temperature to that at near-zero temperature ($T=\SI{1e4}{\K}$) provided an estimate of the plasma degeneracy. 

\begin{figure}
    \centering
    \includegraphics{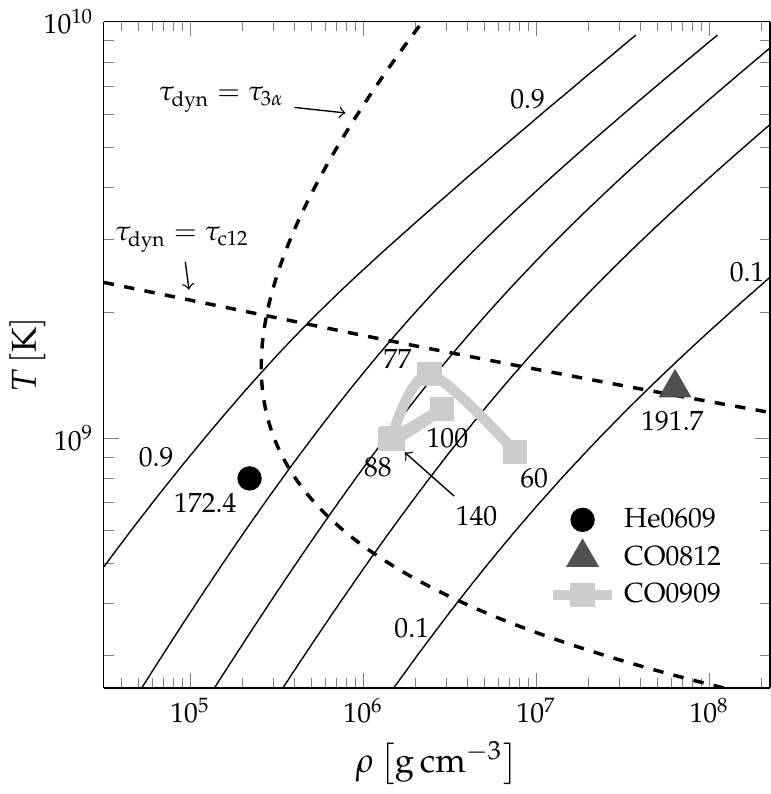}
    \caption{
        Characteristics of the merger models in the density-temperature plane. The data are shown for the hottest mesh cell in the CO0909 model at select times (light grey line, squares). The conditions at the detonation ignition point are indicated with solid circles and triangles for the He0609 and CO0812, respectively.  The labels attached to data points provide corresponding simulation times. Solid contour lines show the fractional contribution of thermal pressure to the total plasma pressure, ranging from 0.1 to 0.9, and are used to indicate the degree of degeneracy. Dashed lines correspond to the regions where the plasma ignition time is equal to the local dynamical timescale for carbon-carbon fusion and triple-alpha reactions. \label{f:tempDensTrajectory}
    }
\end{figure}

A summary of the above results, along with the trajectories of the hottest points in our merger simulations, are shown in Figure~\ref{f:tempDensTrajectory}.
%
%
%
In the figure, the plasma ignition times are shown with dashed lines, while the results of our plasma degeneracy study are drawn with thin solid lines. The labels attached to the thin solid lines denote the ratio of the thermal pressure to the total pressure. The density-temperature trajectory through time of the hottest point in the CO0909 model is shown with a light grey line for select times; the conditions at the detonation ignition points in the CO0812 and He0609 models are marked with triangles and solid circles, respectively.

Furthermore, the explosively-burning plasma may not produce a successful detonation due to curvature effects. This problem has been discussed in detail by a number of authors \citep[][and references therein]{sharpe+01,dunkley+2013}, with the key conclusion being that a sustained detonation can only be produced by an explosively-burning region of a certain critical minimum size. Assuming spherical symmetry, \citet{dursi+06} estimated the dependence of such critical sizes as a function of plasma density and composition. The dependence of the detonation outcomes on curvature effects thus requires numerical models to have spatial resolution sufficient to resolve the critical sizes of detonation ignition kernels \citep{dunkley+2013}. Table~\ref{t:resolution} provides estimated radii of detonation ignition kernels in our models. In our estimation we used Eq.~11 of \citet{dursi+06} and conditions characterizing stellar plasma just prior to its ignition. We defer discussion of the role of numerical resolution on ignition outcomes to Section~\ref{s:meshResolution}.
\subsection{Systematic sensitivity studies} \label{s:sensitivityStudies}

\subsubsection{Orbital separation and inspiral rate}\label{s:orbitalParams}
There exists a certain amount of uncertainty in regard to the final stages of orbital evolution prior to mass transfer in white dwarf binary systems. This is chiefly due to the fact that their orbital evolution is driven by the loss of orbital angular momentum. \citet{webbink+84} first considered the loss of angular momentum due to emission of gravitational waves as the primary mechanism leading to the white dwarf merger.  Simple analytic estimates show that the rate of inspiral is very small compared to the orbital period, even for massive white dwarf pairs. Also, gravitational wave emission is typically not modelled in merger simulations, because the signal, and so potential for observational detection, is relatively low. For these reasons, the orbital evolution just prior to merger is simply parameterized. \citet{even+2009} discussed the importance of constructing accurate initial conditions for white dwarf merger studies using a grid-based code. They used an approximate equation of state of degenerate matter and employed a variant of Hachisu's SCF method to obtain sequences of binary white dwarf models at the onset of mass transfer. (In passing, we note that their models were essentially identical to those used as initial conditions in our study.) 

\begin{figure*}
    \setlength\figureheight{\columnwidth}\ignorespaces 
    \setlength\figurewidth{1.66667\columnwidth}\ignorespaces
    \centering
    \includegraphics{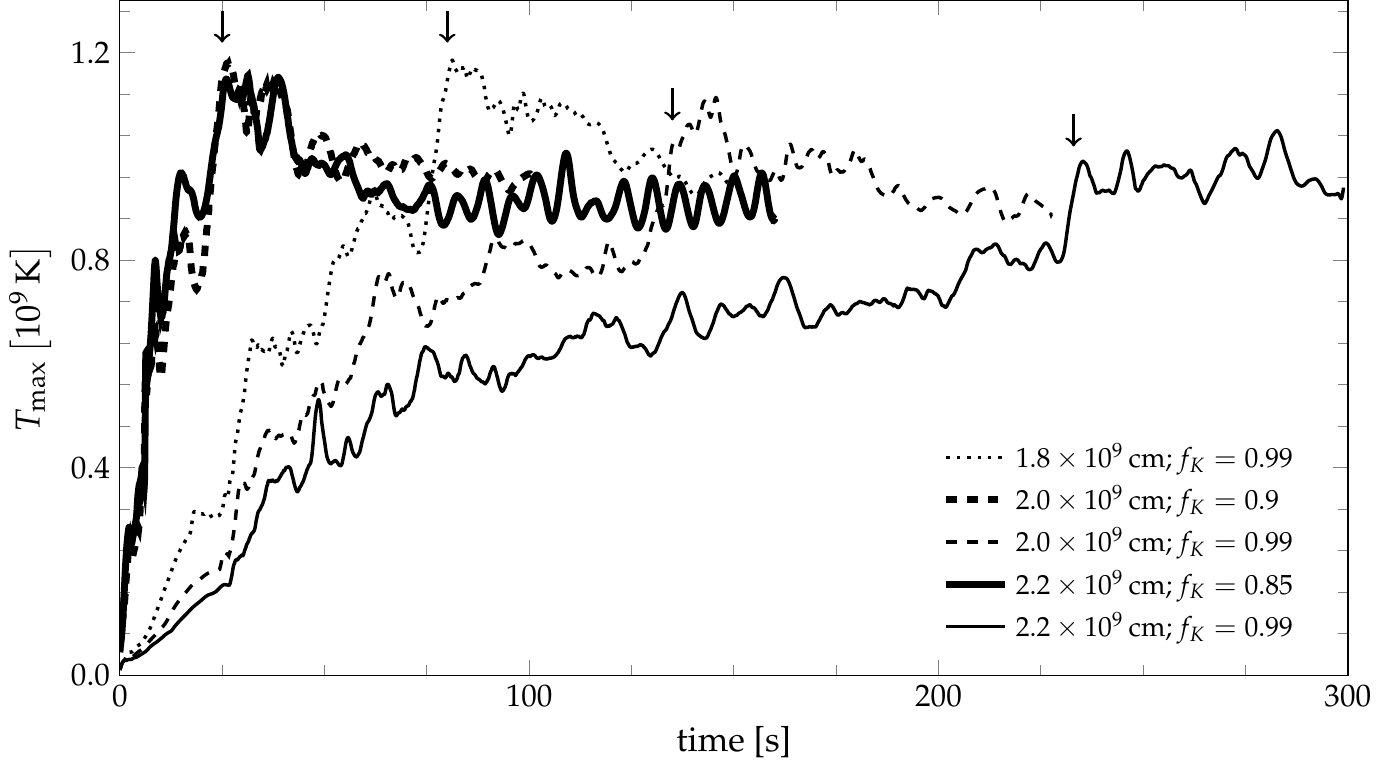}
    \caption
    {
        Evolution of the maximum temperature for various initial separations in the CO0909 model family. Note that in all five cases shown here, the temperature increases at early times due to tidal heating. Arrows indicate the times of accretion shock formation, when the maximum temperature rapidly increases. \label{f:maxTEvolution0909}
    }
\end{figure*}

One of the important aspects of early SPH white dwarf merger models was the use of simplified initial configurations, with mass distributions calculated using a point-mass gravitational potential. As we remarked in Section~\ref{s:selfConsistentField}, \citet{dan+11} proposed a relaxation method to address this deficiency. These authors found that simplified, unrelaxed SPH models are prone to produce unrealistic mass transfer rates. In consequence, the densities and temperatures found in the hot, shocked material will be exaggerated, resulting in artificially short ignition times, enhanced burning, and possibly resulting in spurious ignitions. Some characteristics of these types of models can be found in the work of \citet{pakmor2010}, \citet{tanikawa+2015}, and \citet{sato+2015}. In particular, \citet{pakmor2010} considered a series of white dwarf binaries with mass ratio close to one and relatively massive ($\sim\SI{0.9}{\solarmass}$) components. In the models with fast inspiral rates, he found high densities and  temperatures strongly suggestive of explosive burning. This shows that one can potentially induce detonations by specifically tuning the initial orbital parameters. Therefore, and because, as we noted earlier, direct modeling of the inspiral phase is computationally infeasible, it is important to assess the role of the parameterization used to describe this phase. To this end, we performed a set of merger simulations, systematically varying the initial orbital separation and orbital velocity relative to the Keplerian velocity for every model family. For each simulation, we recorded the time evolution of the maximum temperature. Table~\ref{t:orbitalParam}
\ctable[
    cap = Binary Orbital Parameters,
    caption = {Dependence of the maximum model temperature evolution and the estimated rate of tidal heating on the initial binary separation and orbital frequency.},
    label = t:orbitalParam,
]{cclcr}{
    \tnote{$f_K=\omega /\omega_K$, where $\omega_{K}$ is the Keplerian orbital frequency at the initial time.}
}
{\FL
Model   & $d_{i,9}$                 & $f_K$\tmark  & $\dot{T}_{\mathrm{tidal}}$  & $t(T_\mathrm{max})$     \NN
        & $[\SI{e9}{\cm}]$          &                                   & $[\SI{e7}{\K/\s}]$  & $[\si{\s}]$                     \ML
HE0609  & 2.6                       & 0.99                              & 1.4                     & 72                          \NN
HE0609  & 2.9                       & 0.999                             & 0.5                     & 173                         \NN
CO0812  & 2.2                       & 0.99                              & 2.0                     & 95                          \NN
CO0812  & 2.5                       & 0.999                             & 0.9                     & 192                         \NN
CO0909  & 1.8                       & 0.99                              & 1.9                     & 77                          \NN 
CO0909  & 2.0                       & 0.9                               & 3.2                     & 27                          \NN
CO0909  & 2.0                       & 0.99                              & 1.4                     & 131                         \NN
CO0909  & 2.2                       & 0.85                              & 2.7                     & 31                          \NN
CO0909  & 2.2                       & 0.99                              & 1.0                     & 246                         \NN
CO0909  & 2.6                       & 0.99                              & 0.7                     & 250                         \LL 
}                                    
presents model parameters used in our sensitivity study, along with the average rate of maximum temperature increase, measured over the first orbital period, and the time when the peak temperature was achieved. Because shock heating due to accretion does not occur at early times, the initial evolution of the maximum temperature is due to tidal heating in the cores of the binary system components. Thus, the rate of the maximum temperature increase directly reflects the overall dynamics of the merger. This is confirmed by our estimates of the maximum temperature increase rates. For example, we observed the highest values in models with the tightest initial orbits and the fastest inspiral rates. The two most extreme cases here are models CO0909 with $(d_{i,9},f_K)=(2.0,0.9)$ and $(d_{i,9},f_K)=(2.2,0.85)$. On the opposite extreme, the temperature increased the slowest in the models with the widest initial orbits and smallest inspiral rates. Here, the two most extreme examples are the HE0609 model with $(d_{i,9},f_K)=(2.9,0.999)$ and the CO0909 model with $(d_{i,9},f_K)=(2.6,0.99)$. The variation between these extremes is by a factor of about 6 in our sample. The impact of the initial orbital separation on the maximum temperature evolution is illustrated in Figure~\ref{f:maxTEvolution0909}
for three CO0909 models with different initial orbital separations. In all three cases, the evolution displays a very similar character. Initially, the temperature smoothly increases when the two stars remain nearly stationary on the mesh. However, as soon as advection effects become important and the material becomes perturbed, the maximum temperature begins to display rapid, low-amplitude variations. Conspicuous jumps in temperature shortly before the maximum temperature is reached, indicated with arrows in Figure~\ref{f:maxTEvolution0909}, mark the formation of the accretion shock. It is important to note that as the initial orbital separation increases, the shock forms, and the maximum temperature is obtained later, while the peak temperature decreases. It is clear that model systems with small initial orbital separations produce conditions amenable to explosive burning--possibly detonations. In the case that detonations are not obtained, rapid variations of the maximum temperature continue, while the temperatures begin to gradually decline.

It is interesting to note that similar heating rates can be obtained in models with different orbital separations, provided that the inspiral rate in the model with the wider orbit is suitably higher. This is illustrated in Figure~\ref{f:maxTEvolution0909}, where we show the results obtained for CO0909 models with $(d_{i,9},f_K)=(2.2,0.85)$ and $(d_{i,9},f_K)=(2.0,0.9)$. As discussed in the previous paragraph, these are the two models with the highest rates of tidal heating. In these models, the maximum temperature evolves in an almost identical fashion, reaching similar maximum temperatures, with the accretion shock forming around the same time ($t\approx\SI{24}{\s}$). The effects of varying the inspiral rates are readily visible by comparing the results obtained for the fast-inspiraling $(d_{i,9},f_K)=(2.2,0.85)$ model with the slowly-inspiraling $(d_{i,9},f_K)=(2.2,0.99)$ model. In the latter case, not only is the rate of temperature increase lower by a factor of almost 3, but the shock formation is delayed by over \SI{200}{\s}, or about 5 orbital periods. Given the nonlinear character of the merger process, it is conceivable that the final outcomes of these two models could be potentially very different. This is because the process of tidal disruption of the secondary should be correctly represented in time in order to allow for the gradual development of the turbulent boundary layer (as observed in systems with unequal-mass components), or the process of entraintment of fast-rotating material from the disrupted secondary (as observed in systems with a mass ratio close to one).

Therefore, as a rule, one should proceed with a conservative choice of initial orbital parameters with well-separated components and slow inspiral rates. A more aggressive choice of initial conditions or other model parameters can be justified only if such conservative, reference solutions are obtained. For example, we lowered the inspiral rates and increased the initial orbital separation from their nominal values in the case of the detonating He0609 and CO0812 models. Such modifications of the initial conditions result in less violent evolution of the system, conceivably providing conditions less conducive to detonations. On the contrary, in the case of the CO0909 models, in which no detonations were observed for a conservative choice of initial conditions, we performed a series of simulations with a progressively more aggressive choice of initial orbital parameters. Although the sensitivity studies are potentially the most time-consuming part of simulations, they provide invaluable insights into the nature of modelled systems, and increase confidence in simulation outcomes.

\subsubsection{Self-heating due to nuclear burning} \label{s:selfHeating}
%
One factor contributing to the thermodynamic evolution of the merger, unrelated to the choice of orbital parameters, is the heating of plasma due to nuclear burning. This process has the potential to significantly modify plasma temperature, especially under degenerate conditions. Yet, the effects of nuclear burning were frequently not accounted for in past studies \citep[see, e.g.,][]{dan+11,moll+2014,sato+2015,tanikawa+2015}. To quantify this effect, we performed additional simulations of the CO0812 model without nuclear burning, and at a higher resolution. 

\begin{figure}
    \centering
    \includegraphics{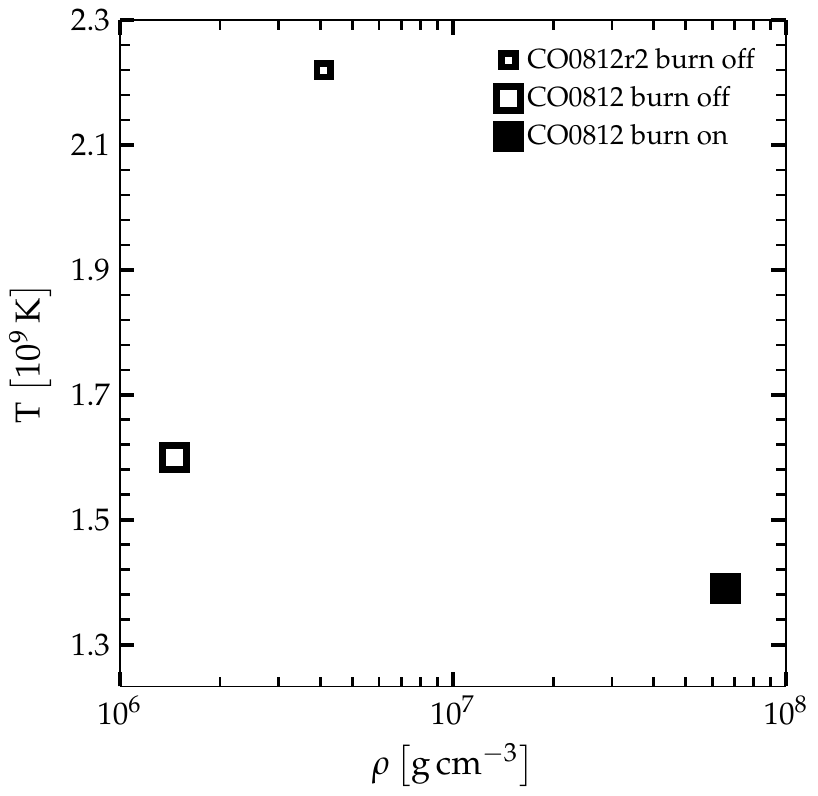}
    \caption
    {
        Density and temperature dependence on self-heating due to nuclear burning and model resolution for select CO0812 model realizations. For the model with self-heating, these are the conditions immediately prior to detonation ignition. For the remaining two models, the data corresponds to the conditions at the times and locations where ignition was most probable, as determined by analysis using ignition times. Note that the maximum temperatures in models without self-heating are achieved at low densities corresponding to the boundary layer, while in the model with self-heating accounted for, the maximum temperature is obtained in the dense core of the primary. See text for a detailed discussion. \label{f:burningComp}
    }
\end{figure}

Figure~\ref{f:burningComp}
%
%
shows the density and temperature at the time that either the minimum ignition time is reached (in models without burning, shown with open squares), or just prior to detonation (in the model with the nuclear heating taken into account, shown with a solid square). With no pre-heating due to nuclear burning, the conditions most favorable for detonations (shortest ignition times) are found in the boundary layer at relatively low densities, essentially precluding a detonation due to weak plasma degeneracy. Furthermore, increasing the resolution to \SI{2}{\km} did not qualitatively change the conditions (the relevant results are shown with a small open square in Figure~\ref{f:burningComp}). However, in the model with thermonuclear preheating, the plasma ignites under degenerate conditions at high densities in the region where no shock-heated material was present and temperature evolution was controlled by gravitational heating and nuclear burning. Since the gravitational heating was also included in the models without burning, we conclude that it is pre-heating that plays a decisive role in detonating the dense core material of the primary. 

Although we are unable to separate the effects of plasma heating from gravitational heating, the contribution to heating due to nuclear burning of the plasma parcel that produced the detonation is reflected in the decrease of its carbon abundance to 0.42 just prior to ignition. We should also note that a compositionally-realistic model of the \SI{1.2}{\solarmass} primary would have a significantly lower central carbon abundance, which in turn would result in a lower rate of self-heating and increased ignition time. We defer discussion of this effect to a future work.  
\subsubsection{The role of numerics} 
\paragraph{Mesh resolution} \label{s:meshResolution}
As we mentioned in Section~\ref{s:parameterSpace}, adequately resolving the large-scale flow features (accretion stream, accretion shock, boundary layer) requires an effective mesh resolution of at least \SI{128}{\km}. On what we consider as intermediate scales, the evolution is dominated by fluid flow instabilities such as the Kelvin-Helmoltz instability. Inspection of our models shows that the typical KHI wavelength at the base of the boundary layer (the interface between the bulk of the primary and the shocked, accreted material) is on the order of 500 to \num{1000} kilometers. Since the PPM hydro solver that we used typically requires about 10 cells to adequately resolve a single wavelength perturbation, a resolution of at least \SI{64}{\km} is needed to correctly capture mixing due to KHI. Of course, still smaller resolution is needed to describe secondary instabilities developing in the boundary layer and mixing in the bulk of the primary components (for example, in our CO0812 and CO0909 models). We note that the reference resolution in our weakly-burning models (CO0609 and CO0909) was \SI{32}{\km}, but we also obtained a sample of models with an effective resolution of \SI{16}{\km}. We used an effective mesh resolution of \SI{16}{\km} in the case of the carbon-detonating CO0812 model, while we studied the interaction between the helium detonation front and the carbon-rich layers of the primary in the He0609 model with a resolution as fine as \SI{2}{\km}. 

Table~\ref{t:resolution}
\ctable[
    cap = {Model resolutions and critical sizes of detonation ignition kernels},
    caption = {Estimated critical sizes of detonation ignition kernels ($D_{\mathrm{k}}$) and maximum effective mesh resolution ($\Delta x_{\mathrm{ef}\mathrm{f}}$) in the merger models.},
    label = {t:resolution},
]{crr}{}
{\FL
model           & $D_{\mathrm{k}}$ [\si{\km}]  & $\Delta x_{\mathrm{ef}\mathrm{f}}$ [\si{\km}]     \ML
He0609          & 13.9                         & 2                                      \NN
CO0609          & --                           & 32                                     \NN
CO0812          & 0.1                          & 16                                     \NN
CO0909          & 4.7                          & 16                                     \LL          
}
lists the highest effective mesh resolutions used in our study, along with the estimated critical diameters of the detonation ignition kernels. We estimated the critical size of the ignition kernel for the CO0812 model using conditions in the ignition region, just prior to ignition. For the He0609 model, the kernel size was calculated from conditions in the carbon-rich layer immediately adjacent to the boundary layer, while for the CO0909 model, the kernel size was calculated from conditions in the region exhibiting the shortest ignition times. We did not estimate the kernel size for the weakly-burning CO0609 model. 

Model configurations similar to our CO0909 model were extensively studied in the past, and produced one of the first detonating white dwarf binary merger models \citep{pakmor+10}. Since we found no evidence for ignition in our \SI{32}{\km} model, contrary to previous studies, we performed an additional simulation with a resolution of \SI{16}{\km}. The evolution of the hottest point in this better-resolved model, as depicted in Figure~\ref{f:tempDensTrajectory}, was \emph{quantitatively} similar to that in the less-resolved model, and thus we did not conduct further mesh resolution studies for the CO0909 model. We conclude that detonations during the early merger stages in such models are unlikely, and the explosions found for these types of models in previous works were a consequence of the specific choice of initial conditions, as we discussed in Section~\ref{s:orbitalParams}. 

Although we were unable to resolve the ignition kernel in the CO0812 model, we performed a sequence of simulations with progressively higher resolutions in order to increase our confidence in finding a detonation. In this series, as the resolution increased from \SI{64}{\km} to \SI{16}{\km}, the detonations were ignited at approximately $t=\SI{209}{\s}, \SI{180}{\s}, \SI{192}{\s}$. We note that, although the ignitions occured at different times, the time difference between detonation formations decreased with increasing resolution, from 29 to \SI{12}{\s}. Even though the performed series of simulations does not provide conclusive evidence for model convergence, as we simply do not resolve the ignition kernel, the observed gradual decrease in the differences between detonation times indicates that the evolution prior to ignition in the detonating region is relatively smooth (see discussion in Section~\ref{s:coreCarbonIgnition}), and suggests that the identified detonation mechanism is robust.

The chaotic (most likely turbulent) nature of the flow in the boundary layer renders the flow in this region more susceptible to resolution effects. In the context of double detonations, adequately capturing the evolution on small scales in the He0609 model proved especially important. As discussed in Section~\ref{s:shockFocusing}, we found no evidence for strong, converging shock waves capable of igniting carbon in the primary's core region. This eliminates the possibility of core ignition. Therefore, we turned our attention to the edge-lit scenario and studied the interaction between the helium detonation front with the carbon-rich primary material immediately adjacent to the boundary layer. In our simulations, this interface region is characterized by large density and compositional gradients, with both the density and carbon abundance increasing toward the base of the boundary layer. From the point of view of carbon ignition, the relevant spatial scale is the carbon ignition kernel diameter, which we estimated to be \SI{13.9}{\km}. We performed a series of simulations with progressively higher resolution, and found no evidence for significant carbon burning, even in the \SI{2}{\km} model, in which the ignition kernel was well-resolved. We conclude that no edge-lit ignition is possible in systems with C/O primaries with masses smaller than \SI{1}{\solarmass}. The edge-lit ignition mechanism remains a possibility in systems with more massive primaries, and clearly merits further study.

\paragraph{Total mass conservation}
As we argued in Section~\ref{s:methods}, it is important that the total mass is preserved on the mesh throughout the course of the simulation in order to correctly describe the evolution of the gravitational potential. We are not aware of any of the previous grid-based studies discussing this important aspect of merger simulations. (The mass is by construction conserved in SPH simulations.) The mass can change (increase or decrease) in a grid-based simulation, even if the underlying hydrodynamics scheme is conservative, due to flow of the material through the boundaries. In such studies, outflow boundary conditions must be used in order to prevent contamination of the interior solution by waves reflected from the boundaries. Then, the only way to avoid mass change in the domain is to extend its boundaries sufficiently. We performed a series of merger simulations with varying sizes of computational domains, and found that the maximum relative change in the total mass in our simulations was \num{-3e-8} provided that the domain was at least $\approx\SI{4.19e11}{\cm}$ in every direction.
\paragraph{Conservation of angular momentum}
Another numerical aspect omitted from the discussion of grid-based white dwarf merger simulations is conservation of angular momentum. This is of particular concern in situations where angular momentum transport plays a key role in the evolution, such as the disc-planet interaction and the evolution of proto-planetary discs \citep[see, e.g.,][and references therein]{valborro+2006}. We found that angular momentum was well-conserved in our simulations. The relative rate of change of the total angular momentum per orbital period varied between \num{4.5e-4} and \num{7.3e-4} for the CO0812 and He0609 models, respectively. The relative change of angular momentum per period was the quantity used by \citet{rosswog+2015} in his discussion of the impact of the nonconservation of angular momentum on the white dwarf merger dynamics. He argued that even modest losses in angular momentum of 0.5 percent per orbit will decrease the merger timescale by a factor of three. Changes in angular momentum in our simulations were substantially smaller than this value, and in the worst cases not dissimilar from the reference value of 0.01 percent used by Rosswog.
%
%
%
%
%
%
\subsubsection{Nuclear network} \label{s:sensitivityNuclearNetwork}
As we demonstrated in Section~\ref{s:selfHeating}, self-heating due to nuclear burning plays an important role in merger simulations. Therefore, the accuracy involved in computing the nuclear energy release can have important implications for the simulation outcomes. \citet{timmes99} demonstrated how the amount of energy produced by nuclear burning depends on the network size. He showed that small alpha-type networks are capable of producing relatively accurate results as compared to large networks. This was an encouraging result that enabled widespread use of reduced nuclear networks in multidimensional hydrodynamic simulations. In the context of the present study, with the white dwarf composition being dominated by helium, carbon, and oxygen, such reduced networks are appropriate to use, and expected to produce nuclear energy with acceptable accuracy.  However, because of the aforementioned sensitivity of detonation ignition to thermonuclear self-heating, it is worthwhile to assess the variations in the model evolution due to the specific choice of nuclear network. To explore this problem, we used a 19-isotope alpha network available in the \textsc{FLASH} code, and a 21-isotope alpha network by \citet{ftimmeswebsite} that we added to the \textsc{Proteus} code. The 21-isotope network differed from the 19-isotope network in that its ODE integrator continuously updated information about nuclear screening and cooling due to neutrinos (in the case of the 19-isotope network, the information about screening and neutrino emmission was calculated only at the beginning of the integration process). We found no qualitative differences between simulations obtained with different networks in the case of the CO0812 model, while we observed slight variations in the timing of the helium detonation in the He0609 models. This is not surprising, given that we are primarily concerned with the ignition conditions at low densities, where the effects due to plasma screening and neutrino emmission are essentially negligible.

We conclude that the key results of our merger simulations are insensitive to the choice of the alpha network and the way that additional physics, such as screening and neutrino emmission, are taken into account. This conclusion does not preclude the possibility that different outcomes may be obtained in models with more realistic chemical compositions \citep[see, e.g.,][]{shen+2014}.
\subsection{He0609: failure of carbon ignition} \label{s:detonationsHeliumBL}
As mentioned in the Introduction, one of the most popular theories of SNe Ia occurring from white dwarf mergers is the double detonation scenario, in which a helium detonation leads to either carbon ignition at the helium-carbon interface, or the converging shock ignites carbon in the core \citep[][and references therein]{livne90}. As stressed originally by \citet{livne+90} (see also \citet{livne90}), multidimensional effects, details of the helium ignition, and numerical effects may play a crucial role in igniting carbon. The fact that previous studies frequently used idealized assumptions (assumed symmetry of the problem, static initial model), and no carbon detonations were produced in our self-consistent model, motivated a study of the evolution of the helium detonation and its interaction with the carbon-rich primary at much higher resolutions. We present the results of this study in the following two sections.
\subsubsection{High resolution study of edge-lighting}
\begin{figure*}

\adjustThreePanels\ignorespaces
\begin{center}
    \begin{tabular}{ccc}
        %
        \includegraphics{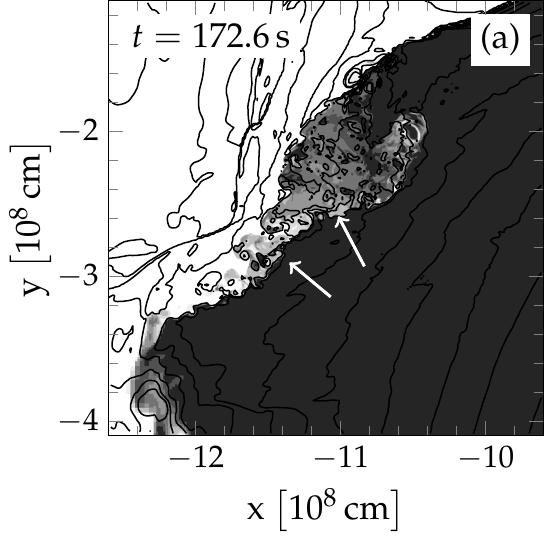}
        &
        \includegraphics{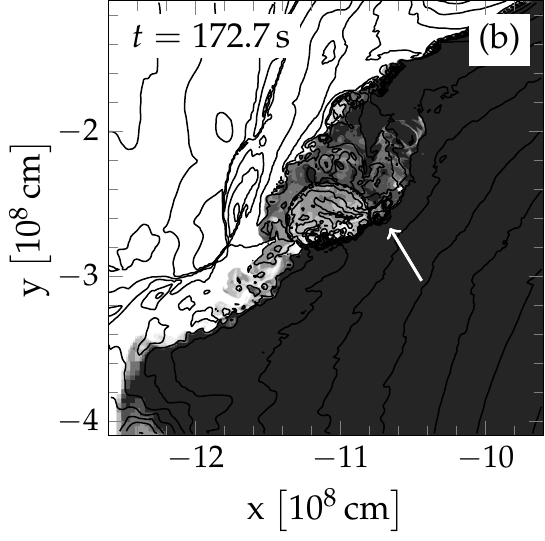}
        &
        \includegraphics{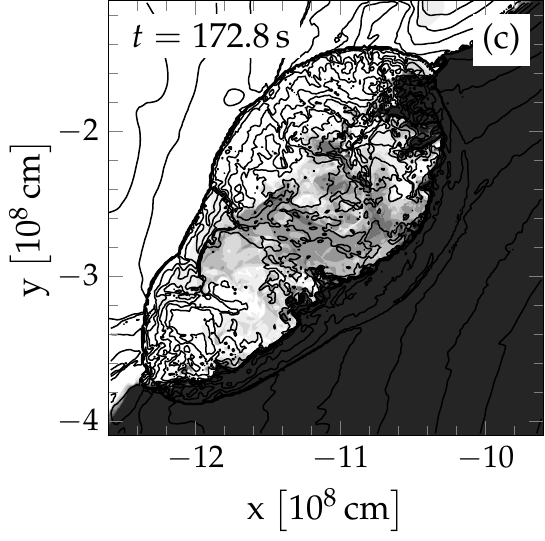} 
    \end{tabular}
%
    \includegraphics{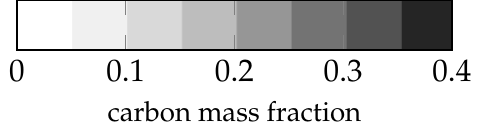}
    \caption
    {
        The ignition at early stages in the evolution of the helium detonation in the He0609 merger model. The distribution of carbon in the model orbital plane is shown with a pseudocolor map in linear scale between 0 and 0.4. The density distribution is shown with 13 contour lines, logarithmically-spaced between \num{1e5} and \SI{4e6}{\g\per\cubic\cm}. The large-scale vortex is occupied with heavily-mixed material, seen in the central region in panel (a). Two arrows visible in this panel indicate one of the secondary vortices (left arrow), and the detonation ignition point (right arrow), while the arrow in panel (b) indicates a vigorously burning pocket of helium. Note that the detonation failed to ignite carbon occupying the trailing section of the vortex, leaving unburned, carbon-rich material protruding into the vortex region. See text for details. \label{f:heDetMechanism}
    }
\end{center}
\end{figure*}

After about \SI{165}{\s}, helium begins vigorously burning in a number of isolated points inside the boundary layer, in the region below the accretion shock. The shock-heated and slowly-burning material was advected downstream and participates in KHI-induced mixing. As we discussed in Section~\ref{He0609results}, the helium detonation was initiated near the base of the boundary layer. To confirm the result obtained in our baseline \SI{64}{\km} model, we performed a series of simulations with progressively-increasing mesh resolution. Figure~\ref{f:heDetMechanism} 
%
%
shows the progress of the detonation from the moment of ignition through the phase of its strong interaction with the adjacent carbon layer. In the figure, the abundance of carbon is shown as a pseudocolor map, while the density is shown with solid contour lines. The region shown in the figure approximately covers the area occupied by one of the large-scale vortices that has a size of $\approx\SI{2000}{\km}$. Immediately prior to ignition (Figure~\ref{f:heDetMechanism}(a)), the vortex contains on average about 50 percent helium, and displays a complex flow pattern with occasional channels and pockets of nearly pure helium. These observed variations in composition on small scales are due to Kelvin-Helmholtz instabilities operating on scales of around $\approx\SI{200}{\km}$, that developed during the early stages of the boundary layer evolution. These instabilities made the interface region ``porous'', with essentially pure helium trapped inside the vortex cores. Several of such helium-rich pockets can be seen at the lower edge of the large vortex, although their appearance was strongly affected by the shear induced on the large scale. However, this shear is not present in the region downstream from the large vortex, allowing for a helium-filled small-scale vortex (created at early times) to retain its original structure (this vortex is located near $(x,y)\approx(\num{-1.14e9},\num{-2.9e8})\, \si{\cm}$, as indicated by an arrow in Figure~\ref{f:heDetMechanism}(a)). Over time, as the boundary layer develops and the accretion rate changes, KHI begins to dominate on much larger scales, leading to the formation of the large-scale vortex. At this time, however, the mixing involves the interface disrupted by KHI on smaller scales, as described above. In the process of mixing, this porous interface overturns and folds, causing compression and providing confinement for burning pockets of trapped helium. 

The detonation is ignited at the point $(x,y)\approx(\num{-1.11e9},\num{-2.6e8})\, \si{\cm}$, located close to the interface between the helium-rich boundary layer and dense carbon (marked with an arrow in Figure~\ref{f:heDetMechanism}(a)). We tracked the igniting parcel back in time for about \SI{150}{\ms}, during which time it traveled about \SI{400}{\km} and was advected to the interface from a height of about \SI{300}{\km} above it. The parcel was initially adiabatically compressed by flow perturbations present in the vortex, with the density increasing from \num{2e5} to \SI{2.6e5}{\gram\per\cubic\cm} over a period of about \SI{50}{\ms}; during this time, the pressure also increased by about 30 percent, while the temperature remained nearly constant. Soon after, the heating rate due to nuclear burning began to steadily increase, causing the material to gradually expand. The self-heating occured under isobaric conditions with a pressure of around \SI{1.15e22}{\dyn\per\cm\squared}. The igniting material had a density and temperature of around \SI{2.2e5}{\gram\per\cubic\cm} and \SI{8e8}{\K}, and contained around  60 percent helium. These conditions place the ignition point close to the ignition line in Figure~\ref{f:tempDensTrajectory}, under weakly degenerate conditions (degeneracy pressure accounted for only around 18 percent of the total pressure). (In passing we note that a small extension of the burning front located somewhat deeper along the interface toward the center of the vortex, marked with an arrow in Figure~\ref{f:heDetMechanism}(b), was created by a separate detonating parcel located about \SI{300}{\km} away from the primary detonation point. This parcel, however, was less abundant in helium and of smaller size. Consequently, it did not produce a strong detonation.) 

Figure~\ref{f:heDetMechanism}(b) shows the resulting detonation wave moving through the helium-rich vortex material and beginning to interact with the unmixed, C/O material of the primary. The density in this region is around \SI{6e5}{\gram\per\cubic\cm}, and the carbon fails to ignite as the post-shock temperatures reach only about \SI{9e8}{\K}. The conditions for carbon ignition appear somewhat more favorable in the region located near the trailing edge of the large-scale vortex, near $(x,y)\approx(\num{-1.08e9},\num{-2.2e8})\, \si{\cm}$, in which the carbon abundance increases more gradually. This allows the wave to retain some of its original strength and heat the material to higher temperatures in this region. Thanks to this, the carbon is partially burned, but the burning ceases as soon as the carbon abundance exceeds about 40 percent, corresponding to a helium abundance lower than about 20 percent. In the process, the shock crosses the interface, leaving behind extended, carbon-rich material protruding into the vortex region, conspicuously visible in Figure~\ref{f:heDetMechanism}(c).
\subsubsection{Effects due to shock focusing in the core} \label{s:shockFocusing}
An additional mechanism whereby carbon ignition has been suggested to originate is via the collision (focusing) of shocks generated by explosive helium burning on the surface of the primary. Our simulations, however, do not support this scenario. In our models, the explosive helium burning is a highly unsteady process, due to strong perturbations present in the boundary layer. This prevents shock convergence. Rather, we observed the conically-shaped weak shocks intersecting in a region located around \SI{3000}{\km} from the center of the primary. We estimated the density of this region to be $\approx \SI{9e6}{\g\per\cubic\cm}$, prohibitively low from the perspective of ignition. At the same time, we did not note any significant temperature change in this region. The final impact of the detonations on the core is its decompression by rarefactions that follow the compressive waves. Furthermore, we note that the core of the primary is itself a dynamic region, with the flow within this region driven by the strongly-varying tidal field. In particular, this causes a large-scale dipolar flow of the material through the primary's core. This material is compressed and heated as it moves through the core region. The heating, however, appears only weak, with the resulting temperatures on the order of \SI{3.5e8}{\K}, and well below the ignition temperature. Overall, the findings that follow from our He0609 study are consistent with the results of \citet{guillochon+10} in that double detonations in white dwarf binary systems with massive C/O primaries and helium-rich secondaries are unlikely.
\subsection{CO0812: core carbon ignition} \label{s:coreCarbonIgnition}
\begin{figure}
    \centering
    \includegraphics{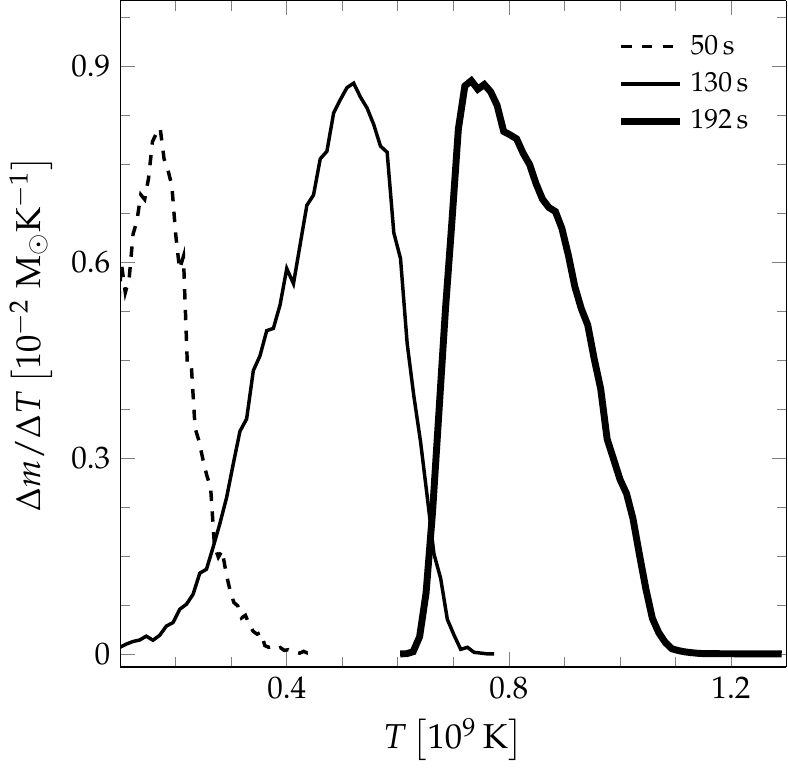}    
    \caption
    {
        Mass distribution as a function of temperature for a hemisphere enclosing \SI{0.2}{\solarmass} centered at the core of the primary for the CO0812 model. The data are shown at three select simulation times, illustrating the progressive heating of the core region via gravitational compression. Note that by $t=\SI{192}{\s}$, the mass distribution has developed a high-temperature tail. The tail is due to self-heating by nuclear burning. \label{f:tempHist0812}
    }
\end{figure}

As described in Section~\ref{s:results0812}, the carbon detonation in the core of the primary was due to compressional heating of the material under strongly degenerate conditions. Figure~\ref{f:tempHist0812}
shows the mass distribution as a function of temperature for a (hemi-spherical, in our simulations) region centered at the core of the primary containing \SI{0.2}{\solarmass}. With time, the maximum of the distribution evolves toward progressively higher temperatures as the primary gains mass. This enables efficient self-heating due to nuclear burning for material residing at temperatures above \SI{1.1e9}{\K}. (The apparent asymmetry of the distribution is an artifact of our data-analysis procedure, which limits the amount of mass involved in the calculations.)

\begin{figure}
    \includegraphics{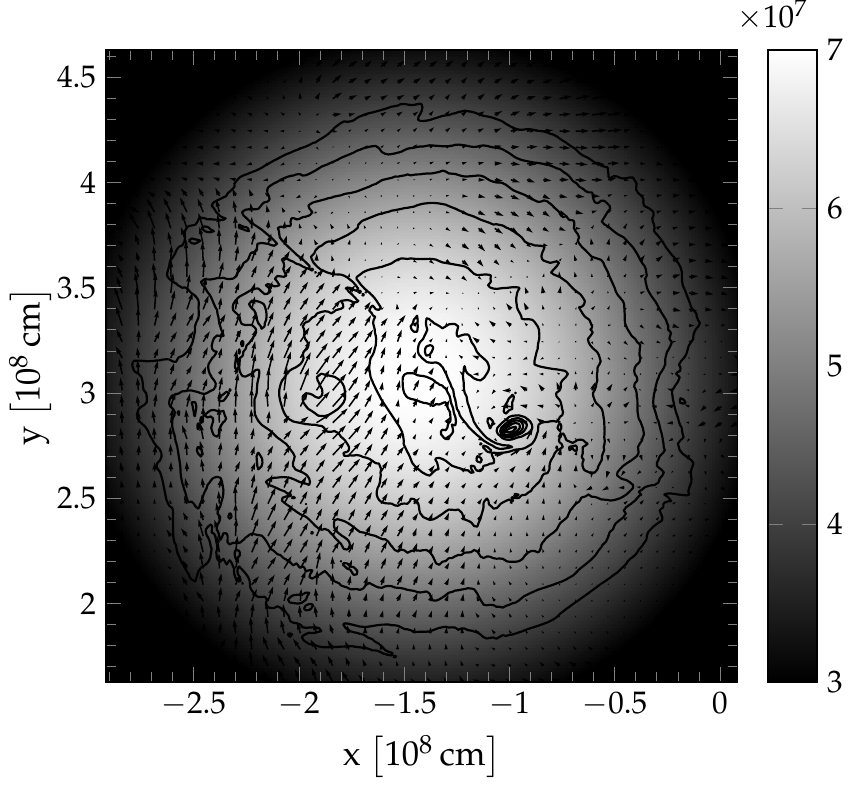}     
    \caption
    {
        Conditions in the core of the CO0812 primary, immediately prior to carbon ignition. The density distribution is shown as a pseudocolor map between \num{3e7} and \SI{7e7}{\g\per\cubic\cm}. There are 10 contours representing the temperature distribution. The temperature contours are logarithmically-spaced from \SI{8e8}{\K} to \SI{1.3e9}{\K}. The arrows depict the direction and speed of the fluid motion. The detonating region can be seen as a closely-spaced set of contours near $(x,y)\approx(\num{-1e8},\num{2.8e8})\, \si{\cm}$. See text for discussion. \label{f:detMech0812}
    }       
\end{figure}

The detonation occurs in the core of a small vortex at a density of about \SI{6.2e7}{\g\per\cubic\cm} and temperature \SI{1.4e9}{\K}, under strongly-degenerate conditions (cf.\ Figure~\ref{f:tempDensTrajectory}). This vortex was produced in the process of slow mixing induced in the core by large-scale flows generated in the outer layers of the primary by accretion and tidal interaction, similar to what we observed in the He0609 model. The resulting flow pattern in the core region immediately prior to the formation of the detonation is shown in Figure~\ref{f:detMech0812}.
The large-scale flow affects roughly half of the core region on the side opposite the disrupted secondary, and the vortex containing the detonation point is located near the interface separating perturbed and unperturbed sections of the core at $(x,y)\approx(\num{-1e8},\num{2.8e8})\, \si{\cm}$. 

Two additional comments are due. First, we note that the self-heating process was aided by the following two phenomena: the vortex appeared to be created with a slightly hotter core than its surroundings, which aided in the process of self-heating due to nuclear burning, and by the fact that the fluid residing in the vortex core was essentially prevented from mixing with the surrounding cooler material. We observed similar effects--enhanced burning and higher temperatures due to self-heating within vortices--in our highly-resolved He0609 model. Second, the observed ignition process is akin to mild ignitions observed in the cores of Chandrasekhar-mass white dwarfs \citep{malone+2014}, and one may expect a competition between the self-heating and thermal conduction in the hotspots associated with vortex cores. However, for the CO0812 core conditions, the thermal conductivity timescale is several orders of magnitude longer than the self-heating timescale. Consequently, we found no qualitative differences in the results of an additional simulation with thermal conductivity enabled.
\subsection{Nucleosynthetic yields and average post-merger structure}
\begin{figure*}
\setlength{\tabcolsep}{1pt}\ignorespaces
\begin{center}
    \begin{tabular}{rrr}
                \multicolumn{1}{c}{\hskip 5em He0609}
                &
                \multicolumn{1}{c}{\hskip 3.5em CO0812}
                &
                \multicolumn{1}{c}{\hskip 3.5em CO0909}
                \\
                \raisebox{-.3\figureheight}{\includegraphics{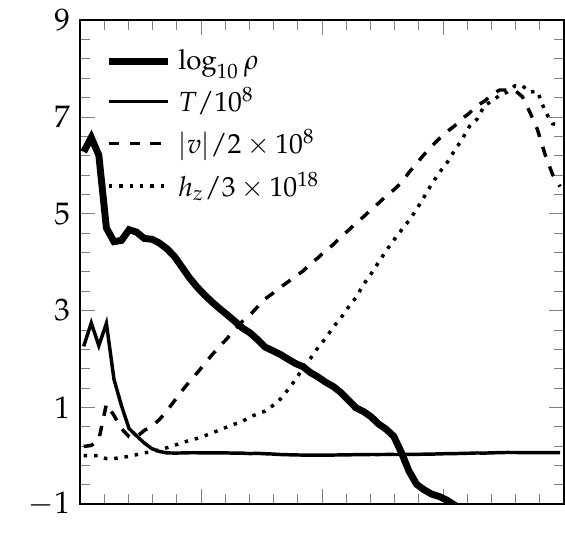}}
                &
                %
                \raisebox{-.3\figureheight}{\includegraphics{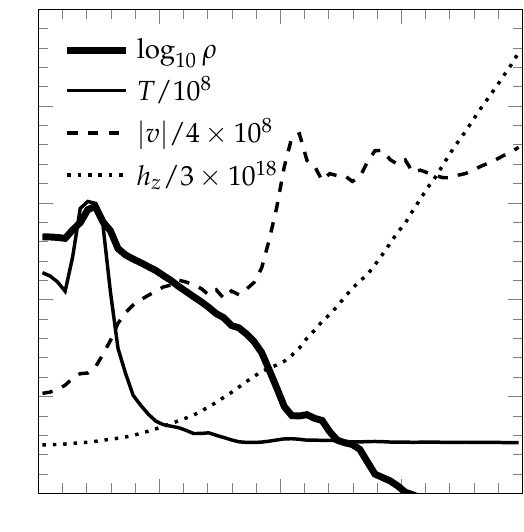}}
                &
                \raisebox{-.3\figureheight}{\includegraphics{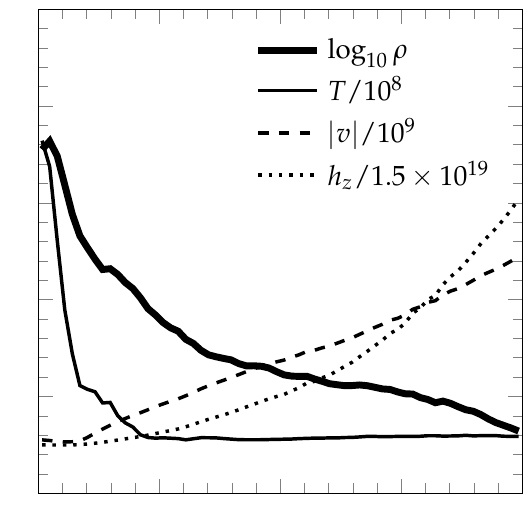}}                %
                \\
                %
                \raisebox{-.3\figureheight}{\includegraphics{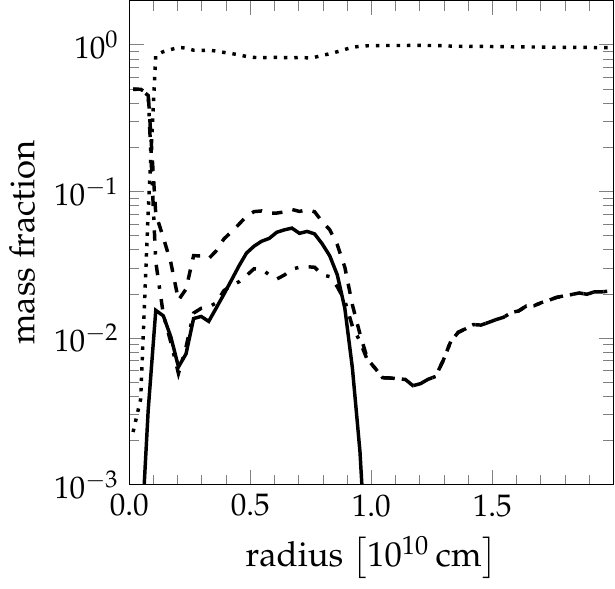}}                %
                &
                %
                \raisebox{-.3\figureheight}{\includegraphics{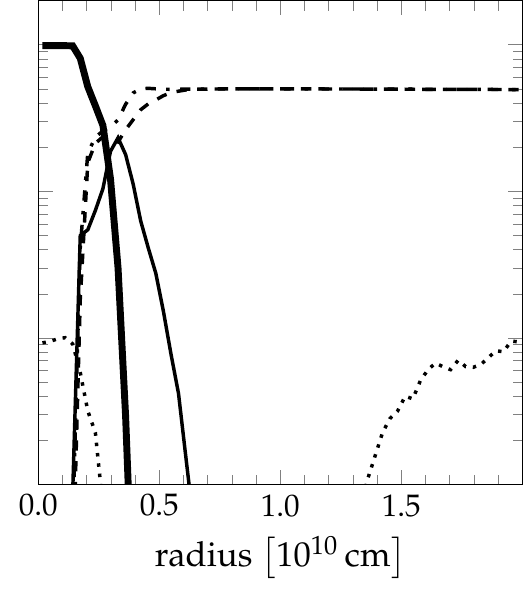}}                %
                &
                \raisebox{-.3\figureheight}{\includegraphics{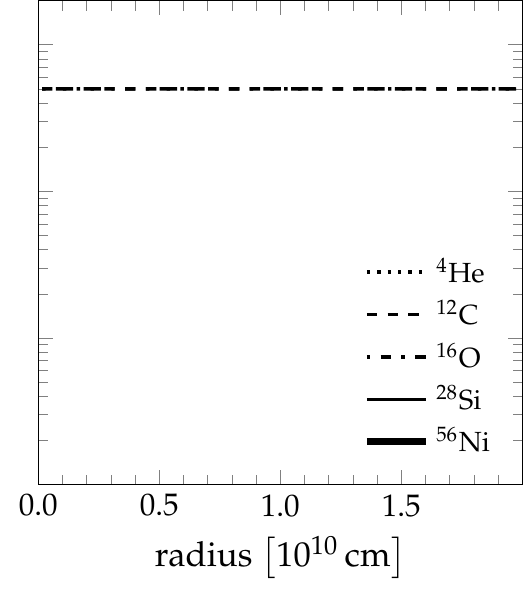}}
    \end{tabular}
    \caption
    {
        Average radial profiles of merger models at the final times. The data for He0609, CO0812, and CO0909 are shown in the left, middle, and right columns of panels, respectively. The average radial profiles of hydrodynamic state variables are shown on the top row. The plotted quantities are density, $\rho \left(\si{\g\per\cubic\cm}\right)$, temperature, $T \left(\si{\K}\right)$, velocity magnitude, $\lvert v \rvert \left(\si{\cm\per\s}\right)$, and the $z$ component of the specific angular momentum, $h_z \left(\si{\cm\squared\per\s}\right)$. The velocity and specific angular momentum are measured in the corotating frame of reference. The average radial compositional profiles for major isotopes are shown in the bottom row. All profiles were calculated using the system's center of mass as the origin, and were suitably averaged over spherical shells at given radii. Note that the scaling of variables and size of the region shown change between panels. See text for discussion. \label{f:finalCond}
    }        
\end{center}
\end{figure*}

Figure~\ref{f:finalCond}
%
shows the average radial profiles of hydrodynamic state variables and composition for each of the binary scenarios. The profiles were obtained at the final simulation times listed in Table~\ref{t:abund}.
\ctable[
    cap = Isotopic abundances,
    caption = {Isotopic abundances, nuclear energy released, and explosion energies for each merger model at the final time. Abundances are given in units of solar mass and energies are given in units of $10^{51}$ ergs.} ,
    label = t:abund,
    star,
]{llllll}{}
{\FL
model  & time    & IME               & IGE          & $E_{\mathrm{nuc}}$ & $E_{\mathrm{exp}}$    \NN
       & [s]   & [\si{\solarmass}]               & [\si{\solarmass}]          & [\SI{e51}{\erg}] & [\SI{e51}{\erg}]    \ML
He0609 & 200        & $2.4\e{-2}$       & $3.2\e{-4}$  & $\num{3.9e-2}$     & --                    \NN
CO0812 & 195        & 0.35              & 0.87         & 2.2                & 1.6                   \NN
CO0909 & 140        & $2.5\e{-6}$       & $\approx 0$  & $\num{1.1e-6}$     & --                    \LL          
}
The radial profiles were obtained by averaging the original simulation data over spherical shells using the system's center of mass as the point of origin. In the figure, the density, temperature, material speed, and specific angular momentum are shown in the top row, while the bottom row shows the run of isotopic abundances. We stopped our simulations when either the abundances did not show significant evolution (for the CO0812 model), or the system's state precluded any chance of a carbon detonation (in the remaining two cases).  

The average profiles in the He0609 model reflect the composite structure of the merger. The average densities in the core are above \SI{e6}{\g\per\cubic\cm} (the density drop at the center is due to our choice of coordinate system, originating at the center of mass, which is displaced from the point of maximum density in the primary). The temperatures in the core are around \SI{3e8}{\K}, and higher than the initial temperatures due to accretion heating. The dense core has a radius of approximately \SI{9e8}{\cm}, which is surrounded by the boundary layer. The layer has an average density of \SI{4e4}{\g\per\cubic\cm}, with a peak temperature of about \SI{3.8e8}{\K} at the base. The highest velocities can be found also at the base of the boundary layer, which are due to the rapid rotation of the accreted material. The core rotates differentially, with the specific angular momentum rising almost linearly from the center. The rotation rate increases rapidly in the transition region between the core and the boundary layer. It reaches a peak value of about \SI{2.2e17}{\cm\squared\per\s} at a radius of approximately \SI{e9}{\cm}, which corresponds to an average rotational velocity of slightly over \SI{2000}{\km\per\s}. The high-density merger core is composed chiefly of the original C/O primary material, with only a small admixture of helium (see the first panel in the second row in Figure~\ref{f:finalCond}). The boundary layer is composed of helium with an admixture of carbon and traces of oxygen and silicon. The abundances of the latter three elements are well-correlated for radii above around \SI{1.4e9}{\cm}, hinting at nuclear burning as their origin.

The average CO0812 model structure significantly differs from the previously-discussed model, due to the energetic explosion taking place in this model. The average density is nearly constant in the central ejecta region, and begins to decrease around \SI{1.2e9}{\cm}, marking the inner edge of the boundary layer material. The boundary layer is significantly hotter ($T\approx\SI{5e8}{\K}$) than the core (temperatures between \num{3e8} and \SI{4e8}{\K}). The hot boundary layer extends to about \SI{3e9}{\cm}. Both the density and temperature show a power-law dependence of roughly $r^{-3}$ outside the shocked boundary layer region, consistent with the constant-pressure of the post-shock region. The temperatures in the shocked circum-binary medium are much lower. At the final time, the expansion of the ejecta core reaches maximum velocities of approximately \SI{5200}{\km\per\s}. The core is composed of iron-group elements (IGE), dominated by radioactive \isotope[56]{Ni} with only a small (about 1 percent) amount of helium. The expansion velocity rather rapidly increases inside the shocked boundary layer, up to about \SI{10000}{\km\per\s}. The composition in this region is rather complex, with its inner portion composed chiefly of \isotope[56]{Ni}, with a trace amount of helium, and the outer portion composed chiefly of unburned carbon and oxygen with up to about 25 percent silicon. The total amounts of nickel and silicon synthesized in this model are about 0.86 and 0.17 solar mass, respectively. The silicon occupies a broad range of velocity space, starting with approximately \SI{5000}{\km\per\s} (at the base of the shocked boundary layer) and extending to peak observed velocities. The ejecta is differentially rotating, with the rate progressively increasing with radius. 

We estimated the typical variations in density in our model by calculating the standard deviations of density from the angle-averaged mean density as a function of radius. The data were obtained for 4096 radial rays equally distributed over the half-sphere containing the bulk of the merger material. The density variations so obtained steadily increase with radius from the center and reach a factor of about 2 at the edge of the nickel-rich core ($r\approx\SI{2e9}{\cm}$). The density fluctuations at this level persist through the boundary layer region dominated by intermediate mass elements ($r<\SI{6e9}{\cm}$), while very large variations (by a factor of 6-8) are observed outside the boundary layer. When the analysis is restricted to the equatorial region (within $30^\circ$ of the equatorial plane), the variations within the nickel-rich core are on the order of 70 percent, and reach a factor of 4-6 in the boundary layer. In contrast, when considering regions within $30^\circ$ of the rotation axis, the variations in the nickel core and the silicon-rich boundary layer do not exceed 10 and 50  percent, respectively. In passing, we note that our estimates of nucleosynthetic yields are sensitive to details of the computational model, such as mesh resolution, choice of nuclear network, and the way nuclear burning is coupled to hydrodynamics. For example, the amount of nickel synthesized in our \SI{64}{\km} resolution model was about 25 percent lower. Therefore, the presented yields should be considered preliminary and obtaining more accurate values would require using a particle post-processing technique \citep[see, e.g.,][]{travaglio+2004}. The fastest-moving material found in our model has a velocity of $\approx \SI{13000}{\km\per\s}$ and is located close behind the supernova shock. At this time, the forward supernova shock has an average radius of approximately \SI{1.05e10}{\cm}.

Finally, the CO0909 model appears to have, on average, the least complicated structure. This is primarily because no explosive burning was observed in this model, and burning was confined to the densest and hottest core part of the merger. This region extends to approximately \SI{5e8}{\cm} from the merger's center, and its outer edge can be defined by a rapid drop in density (seen in the first row in Figure~\ref{f:finalCond}). Two distinct regions can be identified outside the core. First, a relatively slowly, differentially-rotating dense envelope that extends up to a radius of approximately \SI{1.6e9}{\cm}, inside which the average temperature falls from the core temperatures of nearly \SI{7e8}{\K} down to slightly above \SI{1e8}{\K}. The merger's outermost part is a relatively cold ($T \approx \SI{1e8}{\K}$), tenuous ($\rho \approx \SI{3e3}{\g\per\cubic\cm}$), and fast-rotating atmosphere. Of the three models discussed here, the CO0909 model has the most extended structure. No iron-group elements were synthesized in this model, and the burning only produced a trace amount ($\approx \SI{2.5e-6}{\solarmass}$) of neon.
\subsection{Observational properties of the CO0812 explosion model}
The two properties of our CO0812-based model that can be used to provide a rough estimate of the emerging Type Ia supernova model are the nucleosynthetic yields and ejecta morphology shortly after the explosion was launched. Following \citet{mazzali+2007}, we can estimate that for the given amount of IGE species produced in this model, the observed light curve would be characterized by a decline rate of $\Delta m_{15} (\mathrm{B}) \approx 0.99$ and a peak bolometric luminosity of $\SI{1.72e43}{\erg\per\s}$. These parameters would make the CO0812 supernova relatively bright and declining at a nominal rate, although one should be aware of possible orientation effects due to the asymmetry of our model \citep[see, e.g.,][]{kasen+2007, kasen+2009}.

Of note is the large amount of unburned carbon ($\approx \SI{0.35}{\solarmass}$) and oxygen ($\approx \SI{0.42}{\solarmass}$) in the outer layers of the exploding supernova. Spectral signatures of carbon are often difficult to resolve due to blending, and observations must be made soon after the explosion \citep{parrent+2011, parrent+2012, mazzali+2014} for the carbon lines to be observed. A recent study of observations carried out by \citet{parrent+2011} indicates that carbon in SNe Ia is likely more prevalent than previously thought, and may be present in a majority of SN Ia spectra. For example, high signal-to-noise spectral observations of SN 2011fe provided evidence for the presence of both carbon and oxygen \citep{parrent+2012}. Carbon is seen primarily at early times, and oxygen at high velocities, indicating that both occupy the outer layers \citep{parrent+2012, ruiz-lapeunte+2014}. These characteristics are certainly compatible with our model; however, the large amount of carbon and oxygen seen in our model may be difficult to reconcile with observations.

Finally, as we discussed in Section~\ref{s:results0812}, the exploding model is composed of a nearly-spherical ejecta core surrounded by an envelope elongated along the rotation axis. The aspherical shape of the envelope is due to the presence of the secondary material near the orbital plane, which extends away from the orbital plane to various heights along the perimeter of the remnant core. This nonuniformity of distribution of the secondary material near the orbital plane is reflected in the varying strength and vertical extent of the reverse shock, as seen in Figure~\ref{f:mergerMorph_0812_vertical}(c), in which the reverse shock seen to the left of the core is much stronger, and forms up to greater heights away from the orbital plane. Since polarization measurements of Type Ia supernovae point to a nearly spherical shape of these objects, the type of asymmetry observed in the CO0812 model may produce polarization at a level potentially disagreeing with SN Ia observations, unless the supernova is observed pole-on (close to the axis of rotation). However, to resolve this question, our model would have to be evolved until significantly later times, when the expansion becomes homologous. This is beyond the scope of the present work.

%% file: summary.tex
We performed high-resolution simulations of binary white dwarf mergers using an adaptive mesh refinement grid-based hydrodynamic code in three dimensions, assuming symmetry across the orbital plane. We considered binary systems with a total mass in excess of the Chandrasekhar mass. The binary components were cold, C/O or pure helium white dwarfs. Their initial spherical structure was numerically relaxed to account for tidal effects in the initial conditions in a self-consistent way. We used large computational domains to effectively isolate the system and prevent mass flow across the boundaries of the domain. We accounted for the effects of nuclear burning using an alpha-type nuclear network, and used a co-rotating frame of reference to minimize the effects of numerical diffusion. The self-gravity was calculated using a multi-grid approach which is able to account for strong deformations of material that occur during the merger.

We performed a series of simulations with progressively higher mesh resolution. Our focus was on separating the physics of nuclear burning--in particular, detonation ignition--from numerical effects. We observed no vigorous nuclear burning in the case of C/O systems with a total mass below 2 solar masses (systems with 0.6+0.9 and 0.9+0.9 solar-mass components). We found a robust detonation in the C/O system with 0.8 and 1.2 solar-mass components. Furthermore, in the 0.6+0.9 solar mass model with a helium secondary, a detonation was ignited in the boundary layer. The emerging detonation fronts travelled around the primary component, but failed to ignite its material.

Our major findings can be summarized as follows:

\begin{enumerate} 

\item{
    We found a novel detonation mechanism in C/O mergers with massive primaries. The mechanism relies on the combined action of tidal heating, accretion heating, and self-heating due to nuclear burning. It operates in the primary's core region under strongly degenerate conditions. We found that self-heating played a key role in the ignition. The mechanism appears numerically robust, as the detonation was confirmed in a dedicated series of simulations with progressively higher mesh resolution.
}

\item{
    Our study does not support double-detonation scenarios in the case of a system with a \SI{0.6}{\solarmass} helium secondary and a \SI{0.9}{\solarmass} primary. The helium detonation observed in this case appeared too weak to ignite carbon at the base of the boundary layer (edge of the primary), or in the primary's core via focusing of weak shocks driven into its interior by the helium detonations. 
}

\item{
    We found no explosions in models with a primary of 0.9 solar masses (CO0609, He0609, CO0909). This may indicate the existence of a potential mass limit for producing supernova explosions during the violent phase of the mergers \citep[see, e.g.,][]{dan+14}. We did not aim to identify the specific system configurations (total mass, mass ratio) for which detonations in C/O systems occur, and it should be noted that the explosion mechanism found in our C/O merger model differs from  the mechanisms reported in other studies. 
}

\item{
    We observed the formation of a reverse shock in the exploding $0.8+\SI{1.2}{\solarmass}$ C/O system. The reverse shock forms when the detonation escapes from the primary's interior and collides with the boundary layer. The formation process of the reverse shock in this case is identical to the mechanism operating in core-collapse supernovae during the supernova shock passage through the progenitor's envelope. The reverse shock varies in strength, depending on the amount of material encountered outside the primary.
}

\item{
    The exploding $0.8+\SI{1.2}{\solarmass}$ C/O model system is composed of a dense ejecta core, a reverse shock, and a fast-expanding post-shock region. The core is nearly spherically symmetric, and its expansion is constrained near the orbital plane due to the presence of the material from the disrupted secondary. The region bounded by the reverse shock and the main supernova shock is elongated along the rotation axis, with an aspect ratio of about 2. The expanding structure is compositionally stratified, with the core composed chiefly of iron-group elements, while the reverse shock moves into a layer dominated by intermediate-mass elements and unburned material. The model produced about 0.87 solar mass of iron-group elements and approximately 0.35 solar mass of intermediate-mass elements. Roughly 0.8 solar mass of the C/O mix remained unburned.
}

\item{
    Given the energetics and nucleosynthetic yields, the CO0812 model is expected to produce a bright supernova declining at a nominal rate. The model features a relatively large amount of unburned material near the orbital plane. Furthermore, one may expect the observed properties of our model to vary with angle due to the model asymmetry \citep[see, e.g.,][]{kasen+2007, kasen+2009}.
}

\item{
    We conducted a detailed sensitivity study with respect to various parameters of the computational model. In particular, we found the simulation outcomes to be sensitive to the initial orbital system configuration--specifically the orbital separation and the adopted rate of inspiral. We found no dependence on the specific variant of the nuclear network used, as far as the alpha networks are concerned.
}

\end{enumerate}

The results of our study offer a number of possible future research directions. For example, the novel explosion mechanism should be verified in the case of primary white dwarf models with more realistic compositions. A realistic 1.2 solar mass white dwarf model is expected to have a central abundance of carbon reduced by a factor of perhaps 2 compared to the model considered here, resulting in less efficient self heating and longer ignition times. The assumption of symmetry across the orbital plane should be relaxed, enabling unrestricted flow in the vertical direction. Predictions as to how the simulation outcomes will change in models with no assumed symmetry require additional simulations, due to the highly nonlinear character of the flow dynamics. Also, our models should be evolved until later times. In particular, in the case of the CO0812 model, one should aim to produce a homologously-expanding structure to enable more direct comparison with observations. The substantial amount of unburned material obtained in this case may be difficult to reconcile with observations. Such a comparison would require performing multidimensional radiative transfer calculations, as the expanding structure is asymmetric, with unburned material located mainly near the orbital plane. Finally, the further evolution in our non-exploding models could be studied with stellar evolution codes to predict their ultimate fates. 

%% file: appendix.tex
\section{Computational speedup due to load balancing of nuclear burning} \label{a:lbnbSpeedup}

\begin{figure}
    \centering
    \includegraphics{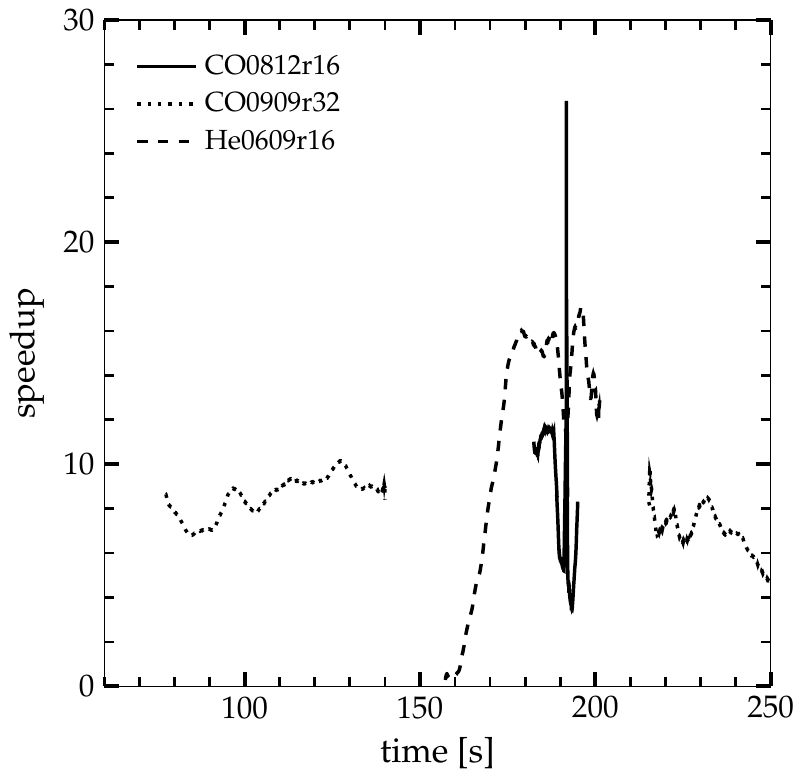}
    \caption
    {
        Computational performance of the nuclear burning load balancer. The estimated speedup gained from using the load balancer is shown for select model realizations. A 19-isotope network is used. For presentation purposes, the data have been smoothed with a boxcar smoother using a window of 1000 hydro steps.  The speedup exceeds a factor of 10 during the phases involving explosive nuclear burning, and reached a factor of 37 and 29 in the CO0812 and He0609 models, respectively. See text for discussion. \label{f:LBNB}
    }
\end{figure}

Balancing the nuclear burning was critical to make our computational study feasible. This is illustrated in Figure~\ref{f:LBNB}
in which we show the time evolution of the speedup obtained thanks to the use of our LBNB load balancer. The data were taken from a subset of our models using a 19-isotope alpha network. The greatest speedup was found in the detonating CO0812 model. Prior to the detonation, the speedup was slightly higher than a factor of 10, and reached a factor of about 37 during the detonation. Similarly, an average speedup of 16 was observed during the explosive helium burning in the He0609 model, with a maximum measured speedup of approximately 29. In the modestly-burning CO0909 model, the gains do not show transient behavior, and allow to speed up calculations on average by a factor of about 8. We would like to note that greater speedups can be expected in the case that larger, more expensive networks are used. For example, in our CO0812 model computed with a 21-isotope alpha-type network \citep{ftimmeswebsite}, not discussed here in detail, we observed speedups by a factor of more than 100 for extended periods of time. Thus, the load balancing for nuclear burning should be considered a true enabler of multidimensional hydrodynamic simulations, if the application requires the use of a large nuclear network. However, even in the case that smaller networks are used, as frequently is the case in exploratory studies, speedups by a factor of a few will improve coverage of the parameter space and quality of the predictions.

%% file: manuscript.bbl
\begin{thebibliography}{}
\makeatletter
\relax
\def\mn@urlcharsother{\let\do\@makeother \do\$\do\&\do\#\do\^\do\_\do\%\do\~}
\def\mn@doi{\begingroup\mn@urlcharsother \@ifnextchar [ {\mn@doi@}
  {\mn@doi@[]}}
\def\mn@doi@[#1]#2{\def\@tempa{#1}\ifx\@tempa\@empty \href
  {http://dx.doi.org/#2} {doi:#2}\else \href {http://dx.doi.org/#2} {#1}\fi
  \endgroup}
\def\mn@eprint#1#2{\mn@eprint@#1:#2::\@nil}
\def\mn@eprint@arXiv#1{\href {http://arxiv.org/abs/#1} {{\tt arXiv:#1}}}
\def\mn@eprint@dblp#1{\href {http://dblp.uni-trier.de/rec/bibtex/#1.xml}
  {dblp:#1}}
\def\mn@eprint@#1:#2:#3:#4\@nil{\def\@tempa {#1}\def\@tempb {#2}\def\@tempc
  {#3}\ifx \@tempc \@empty \let \@tempc \@tempb \let \@tempb \@tempa \fi \ifx
  \@tempb \@empty \def\@tempb {arXiv}\fi \@ifundefined
  {mn@eprint@\@tempb}{\@tempb:\@tempc}{\expandafter \expandafter \csname
  mn@eprint@\@tempb\endcsname \expandafter{\@tempc}}}

\bibitem[\protect\citeauthoryear{{Arnett}}{{Arnett}}{1996}]{Arnett96}
{Arnett} W.~D.,  1996, Supernovae and nucleosynthesis. An investigation of the
  history of matter, from the Big Bang to the present.
Princeton University Press, Princeton

\bibitem[\protect\citeauthoryear{Berger \& Colella}{Berger \&
  Colella}{1989}]{BC89}
Berger M.~J.,  Colella P.,  1989, J. Comput. Phys., 82, 64

\bibitem[\protect\citeauthoryear{Briggs, Henson  \& McCormick}{Briggs
  et~al.}{2000}]{briggs+2000}
Briggs W.~L.,  Henson V.~E.,   McCormick S.~F.,  2000, A Multigrid Tutorial
  (2Nd Ed.).
Society for Industrial and Applied Mathematics, Philadelphia, PA, USA

\bibitem[\protect\citeauthoryear{Childs et~al.,}{Childs
  et~al.}{2012}]{HPV:VisIt}
Childs H.,  et~al., 2012, in , {High Performance Visualization--Enabling
  Extreme-Scale Scientific Insight}.
pp 357--372

\bibitem[\protect\citeauthoryear{Colella \& Woodward}{Colella \&
  Woodward}{1984}]{colella+84}
Colella P.,  Woodward P.~R.,  1984, J. Comput. Phys., 54, 174

\bibitem[\protect\citeauthoryear{{Dan}, {Rosswog}, {Guillochon}  \&
  {Ramirez-Ruiz}}{{Dan} et~al.}{2011}]{dan+11}
{Dan} M.,  {Rosswog} S.,  {Guillochon} J.,   {Ramirez-Ruiz} E.,  2011, \mn@doi
  [\apj] {10.1088/0004-637X/737/2/89}, \href
  {http://adsabs.harvard.edu/abs/2011ApJ...737...89D} {737, 89}

\bibitem[\protect\citeauthoryear{{Dan}, {Rosswog}, {Guillochon}  \&
  {Ramirez-Ruiz}}{{Dan} et~al.}{2012}]{dan+12}
{Dan} M.,  {Rosswog} S.,  {Guillochon} J.,   {Ramirez-Ruiz} E.,  2012, \mn@doi
  [\mnras] {10.1111/j.1365-2966.2012.20794.x}, \href
  {http://adsabs.harvard.edu/abs/2012MNRAS.422.2417D} {422, 2417}

\bibitem[\protect\citeauthoryear{{Dan}, {Rosswog}, {Br{\"u}ggen}  \&
  {Podsiadlowski}}{{Dan} et~al.}{2014}]{dan+14}
{Dan} M.,  {Rosswog} S.,  {Br{\"u}ggen} M.,   {Podsiadlowski} P.,  2014,
  \mn@doi [\mnras] {10.1093/mnras/stt1766}, \href
  {http://adsabs.harvard.edu/abs/2014MNRAS.438...14D} {438, 14}

\bibitem[\protect\citeauthoryear{{Dunkley}, {Sharpe}  \& {Falle}}{{Dunkley}
  et~al.}{2013}]{dunkley+2013}
{Dunkley} S.~D.,  {Sharpe} G.~J.,   {Falle} S.~A.~E.~G.,  2013, \mn@doi
  [\mnras] {10.1093/mnras/stt422}, \href
  {http://adsabs.harvard.edu/abs/2013MNRAS.431.3429D} {431, 3429}

\bibitem[\protect\citeauthoryear{{Dursi} \& {Timmes}}{{Dursi} \&
  {Timmes}}{2006}]{dursi+06}
{Dursi} L.~J.,  {Timmes} F.~X.,  2006, \mn@doi [\apj] {10.1086/500638}, \href
  {http://adsabs.harvard.edu/abs/2006ApJ...641.1071D} {641, 1071}

\bibitem[\protect\citeauthoryear{{Even} \& {Tohline}}{{Even} \&
  {Tohline}}{2009}]{even+2009}
{Even} W.,  {Tohline} J.~E.,  2009, \mn@doi [\apjs]
  {10.1088/0067-0049/184/2/248}, \href
  {http://adsabs.harvard.edu/abs/2009ApJS..184..248E} {184, 248}

\bibitem[\protect\citeauthoryear{{Fink}, {Hillebrandt}  \& {R{\"o}pke}}{{Fink}
  et~al.}{2007}]{fink+2007}
{Fink} M.,  {Hillebrandt} W.,   {R{\"o}pke} F.~K.,  2007, \mn@doi [\aap]
  {10.1051/0004-6361:20078438}, \href
  {http://adsabs.harvard.edu/abs/2007A%26A...476.1133F} {476, 1133}

\bibitem[\protect\citeauthoryear{{Fink}, {R{\"o}pke}, {Hillebrandt},
  {Seitenzahl}, {Sim}  \& {Kromer}}{{Fink} et~al.}{2010}]{fink+10}
{Fink} M.,  {R{\"o}pke} F.~K.,  {Hillebrandt} W.,  {Seitenzahl} I.~R.,  {Sim}
  S.~A.,   {Kromer} M.,  2010, \mn@doi [\aap] {10.1051/0004-6361/200913892},
  \href {http://adsabs.harvard.edu/abs/2010A%26A...514A..53F} {514, A53}

\bibitem[\protect\citeauthoryear{{Fowler} \& {Hoyle}}{{Fowler} \&
  {Hoyle}}{1964}]{Fowler+64}
{Fowler} W.~A.,  {Hoyle} F.,  1964, \mn@doi [\apjs] {10.1086/190103}, \href
  {http://adsabs.harvard.edu/abs/1964ApJS....9..201F} {9, 201}

\bibitem[\protect\citeauthoryear{{Fryxell} et~al.,}{{Fryxell}
  et~al.}{2000}]{Fryxell+00}
{Fryxell} B.,  et~al., 2000, \apjs, 131, 273

\bibitem[\protect\citeauthoryear{{Gilfanov} \& {Bogd{\'a}n}}{{Gilfanov} \&
  {Bogd{\'a}n}}{2010}]{gilfanov+10}
{Gilfanov} M.,  {Bogd{\'a}n} {\'A}.,  2010, \mn@doi [\nat]
  {10.1038/nature08685}, \href
  {http://adsabs.harvard.edu/abs/2010Natur.463..924G} {463, 924}

\bibitem[\protect\citeauthoryear{{Guerrero}, {Garc{\'{\i}}a-Berro}  \&
  {Isern}}{{Guerrero} et~al.}{2004}]{guerrero+04}
{Guerrero} J.,  {Garc{\'{\i}}a-Berro} E.,   {Isern} J.,  2004, \mn@doi [\aap]
  {10.1051/0004-6361:20031504}, \href
  {http://adsabs.harvard.edu/abs/2004A%26A...413..257G} {413, 257}

\bibitem[\protect\citeauthoryear{{Guillochon}, {Dan}, {Ramirez-Ruiz}  \&
  {Rosswog}}{{Guillochon} et~al.}{2010}]{guillochon+10}
{Guillochon} J.,  {Dan} M.,  {Ramirez-Ruiz} E.,   {Rosswog} S.,  2010, \mn@doi
  [\apjl] {10.1088/2041-8205/709/1/L64}, \href
  {http://adsabs.harvard.edu/abs/2010ApJ...709L..64G} {709, L64}

\bibitem[\protect\citeauthoryear{{Hachisu}}{{Hachisu}}{1986}]{hachisu+86}
{Hachisu} I.,  1986, \mn@doi [\apjs] {10.1086/191121}, \href
  {http://adsabs.harvard.edu/abs/1986ApJS...61..479H} {61, 479}

\bibitem[\protect\citeauthoryear{{Holcomb}, {Guillochon}, {De Colle}  \&
  {Ramirez-Ruiz}}{{Holcomb} et~al.}{2013}]{holcomb+2013}
{Holcomb} C.,  {Guillochon} J.,  {De Colle} F.,   {Ramirez-Ruiz} E.,  2013,
  \mn@doi [\apj] {10.1088/0004-637X/771/1/14}, \href
  {http://adsabs.harvard.edu/abs/2013ApJ...771...14H} {771, 14}

\bibitem[\protect\citeauthoryear{{Hoyle} \& {Fowler}}{{Hoyle} \&
  {Fowler}}{1960}]{Hoyle_Fowler}
{Hoyle} F.,  {Fowler} W.~A.,  1960, \mn@doi [\apj] {10.1086/146963}, \href
  {http://adsabs.harvard.edu/abs/1960ApJ...132..565H} {132, 565}

\bibitem[\protect\citeauthoryear{{Hubber}, {Falle}  \& {Goodwin}}{{Hubber}
  et~al.}{2013}]{hubber+2013}
{Hubber} D.~A.,  {Falle} S.~A.~E.~G.,   {Goodwin} S.~P.,  2013, \mn@doi
  [\mnras] {10.1093/mnras/stt509}, \href
  {http://adsabs.harvard.edu/abs/2013MNRAS.432..711H} {432, 711}

\bibitem[\protect\citeauthoryear{{Iben} \& {Tutukov}}{{Iben} \&
  {Tutukov}}{1984}]{iben+84}
{Iben} Jr. I.,  {Tutukov} A.~V.,  1984, \mn@doi [\apjs] {10.1086/190932}, \href
  {http://adsabs.harvard.edu/abs/1984ApJS...54..335I} {54, 335}

\bibitem[\protect\citeauthoryear{{Kapila}, {Schwendeman}, {Quirk}  \&
  {Hawa}}{{Kapila} et~al.}{2002}]{kapila+2002}
{Kapila} A.~K.,  {Schwendeman} D.~W.,  {Quirk} J.~J.,   {Hawa} T.,  2002,
  \mn@doi [Combustion Theory Modelling] {10.1088/1364-7830/6/4/302}, \href
  {http://adsabs.harvard.edu/abs/2002CTM.....6..553K} {6, 553}

\bibitem[\protect\citeauthoryear{{Kasen} \& {Plewa}}{{Kasen} \&
  {Plewa}}{2007}]{kasen+2007}
{Kasen} D.,  {Plewa} T.,  2007, \mn@doi [\apj] {10.1086/516834}, \href
  {http://adsabs.harvard.edu/abs/2007ApJ...662..459K} {662, 459}

\bibitem[\protect\citeauthoryear{{Kasen}, {R{\"o}pke}  \& {Woosley}}{{Kasen}
  et~al.}{2009}]{kasen+2009}
{Kasen} D.,  {R{\"o}pke} F.~K.,   {Woosley} S.~E.,  2009, \mn@doi [\nat]
  {10.1038/nature08256}, \href
  {http://adsabs.harvard.edu/abs/2009Natur.460..869K} {460, 869}

\bibitem[\protect\citeauthoryear{{Kashyap}, {Fisher}, {Garc{\'{\i}}a-Berro},
  {Aznar-Sigu{\'a}n}, {Ji}  \& {Lor{\'e}n-Aguilar}}{{Kashyap}
  et~al.}{2015}]{kashyap+2015}
{Kashyap} R.,  {Fisher} R.,  {Garc{\'{\i}}a-Berro} E.,  {Aznar-Sigu{\'a}n} G.,
  {Ji} S.,   {Lor{\'e}n-Aguilar} P.,  2015, \mn@doi [\apjl]
  {10.1088/2041-8205/800/1/L7}, \href
  {http://adsabs.harvard.edu/abs/2015ApJ...800L...7K} {800, L7}

\bibitem[\protect\citeauthoryear{{Khokhlov} \& {Ergma}}{{Khokhlov} \&
  {Ergma}}{1986}]{khokhlov+86}
{Khokhlov} A.~M.,  {Ergma} E.~V.,  1986, Soviet Astronomy Letters, \href
  {http://adsabs.harvard.edu/abs/1986SvAL...12..152K} {12, 152}

\bibitem[\protect\citeauthoryear{{Kifonidis}, {Plewa}, {Janka}  \&
  {M\"uller}}{{Kifonidis} et~al.}{2003}]{kifonidis+03}
{Kifonidis} K.,  {Plewa} T.,  {Janka} H.-T.,   {M\"uller} E.,  2003, \aap, 408,
  621

\bibitem[\protect\citeauthoryear{{Livne}}{{Livne}}{1990}]{livne90}
{Livne} E.,  1990, \mn@doi [\apjl] {10.1086/185721}, \href
  {http://adsabs.harvard.edu/abs/1990ApJ...354L..53L} {354, L53}

\bibitem[\protect\citeauthoryear{{Livne} \& {Glasner}}{{Livne} \&
  {Glasner}}{1990}]{livne+90}
{Livne} E.,  {Glasner} A.~S.,  1990, \mn@doi [\apj] {10.1086/169189}, \href
  {http://adsabs.harvard.edu/abs/1990ApJ...361..244L} {361, 244}

\bibitem[\protect\citeauthoryear{{Lor{\'e}n-Aguilar}, {Isern}  \&
  {Garc{\'{\i}}a-Berro}}{{Lor{\'e}n-Aguilar} et~al.}{2009}]{loren-aguilar+09}
{Lor{\'e}n-Aguilar} P.,  {Isern} J.,   {Garc{\'{\i}}a-Berro} E.,  2009, \mn@doi
  [\aap] {10.1051/0004-6361/200811060}, \href
  {http://adsabs.harvard.edu/abs/2009A%26A...500.1193L} {500, 1193}

\bibitem[\protect\citeauthoryear{{Lucy}}{{Lucy}}{1977}]{lucy77}
{Lucy} L.~B.,  1977, \mn@doi [\aj] {10.1086/112164}, \href
  {http://adsabs.harvard.edu/abs/1977AJ.....82.1013L} {82, 1013}

\bibitem[\protect\citeauthoryear{{MacNeice}, {Olson}, {Mobarry}, {de
  Fainchtein}  \& {Packer}}{{MacNeice} et~al.}{2000}]{macneice+2000}
{MacNeice} P.,  {Olson} K.~M.,  {Mobarry} C.,  {de Fainchtein} R.,   {Packer}
  C.,  2000, \mn@doi [Computer Physics Communications]
  {10.1016/S0010-4655(99)00501-9}, \href
  {http://adsabs.harvard.edu/abs/2000CoPhC.126..330M} {126, 330}

\bibitem[\protect\citeauthoryear{{Malone}, {Nonaka}, {Woosley}, {Almgren},
  {Bell}, {Dong}  \& {Zingale}}{{Malone} et~al.}{2014}]{malone+2014}
{Malone} C.~M.,  {Nonaka} A.,  {Woosley} S.~E.,  {Almgren} A.~S.,  {Bell}
  J.~B.,  {Dong} S.,   {Zingale} M.,  2014, \mn@doi [\apj]
  {10.1088/0004-637X/782/1/11}, \href
  {http://adsabs.harvard.edu/abs/2014ApJ...782...11M} {782, 11}

\bibitem[\protect\citeauthoryear{{Maoz} \& {Mannucci}}{{Maoz} \&
  {Mannucci}}{2012}]{maoz+12}
{Maoz} D.,  {Mannucci} F.,  2012, \mn@doi [\pasa] {10.1071/AS11052}, \href
  {http://adsabs.harvard.edu/abs/2012PASA...29..447M} {29, 447}

\bibitem[\protect\citeauthoryear{{Mazzali}, {R{\"o}pke}, {Benetti}  \&
  {Hillebrandt}}{{Mazzali} et~al.}{2007}]{mazzali+2007}
{Mazzali} P.~A.,  {R{\"o}pke} F.~K.,  {Benetti} S.,   {Hillebrandt} W.,  2007,
  \mn@doi [Science] {10.1126/science.1136259}, \href
  {http://adsabs.harvard.edu/abs/2007Sci...315..825M} {315, 825}

\bibitem[\protect\citeauthoryear{{Mazzali} et~al.,}{{Mazzali}
  et~al.}{2014}]{mazzali+2014}
{Mazzali} P.~A.,  et~al., 2014, \mn@doi [\mnras] {10.1093/mnras/stu077}, \href
  {http://adsabs.harvard.edu/abs/2014MNRAS.439.1959M} {439, 1959}

\bibitem[\protect\citeauthoryear{{Mochkovitch} \& {Livio}}{{Mochkovitch} \&
  {Livio}}{1989}]{mochkovitch+89}
{Mochkovitch} R.,  {Livio} M.,  1989, \aap, \href
  {http://adsabs.harvard.edu/abs/1989A%26A...209..111M} {209, 111}

\bibitem[\protect\citeauthoryear{{Moll}, {Raskin}, {Kasen}  \&
  {Woosley}}{{Moll} et~al.}{2014}]{moll+2014}
{Moll} R.,  {Raskin} C.,  {Kasen} D.,   {Woosley} S.~E.,  2014, \mn@doi [\apj]
  {10.1088/0004-637X/785/2/105}, \href
  {http://adsabs.harvard.edu/abs/2014ApJ...785..105M} {785, 105}

\bibitem[\protect\citeauthoryear{{Monaghan}}{{Monaghan}}{1992}]{monaghan99}
{Monaghan} J.~J.,  1992, \mn@doi [\araa] {10.1146/annurev.aa.30.090192.002551},
  \href {http://adsabs.harvard.edu/abs/1992ARA%26A..30..543M} {30, 543}

\bibitem[\protect\citeauthoryear{{Nomoto}}{{Nomoto}}{1982}]{nomoto82b}
{Nomoto} K.,  1982, \mn@doi [\apj] {10.1086/160031}, \href
  {http://adsabs.harvard.edu/abs/1982ApJ...257..780N} {257, 780}

\bibitem[\protect\citeauthoryear{{Nugent} et~al.,}{{Nugent}
  et~al.}{2011}]{nugent+2011}
{Nugent} P.~E.,  et~al., 2011, \mn@doi [\nat] {10.1038/nature10644}, \href
  {http://adsabs.harvard.edu/abs/2011Natur.480..344N} {480, 344}

\bibitem[\protect\citeauthoryear{Pakmor}{Pakmor}{2010}]{pakmor2010}
Pakmor R.~M.,  2010, PhD thesis, Technische Universitaet Muenchen

\bibitem[\protect\citeauthoryear{{Pakmor}, {Kromer}, {R{\"o}pke}, {Sim},
  {Ruiter}  \& {Hillebrandt}}{{Pakmor} et~al.}{2010}]{pakmor+10}
{Pakmor} R.,  {Kromer} M.,  {R{\"o}pke} F.~K.,  {Sim} S.~A.,  {Ruiter} A.~J.,
  {Hillebrandt} W.,  2010, \mn@doi [\nat] {10.1038/nature08642}, \href
  {http://adsabs.harvard.edu/abs/2010Natur.463...61P} {463, 61}

\bibitem[\protect\citeauthoryear{{Pakmor}, {Hachinger}, {R{\"o}pke}  \&
  {Hillebrandt}}{{Pakmor} et~al.}{2011}]{pakmor+2011}
{Pakmor} R.,  {Hachinger} S.,  {R{\"o}pke} F.~K.,   {Hillebrandt} W.,  2011,
  \mn@doi [\aap] {10.1051/0004-6361/201015653}, \href
  {http://adsabs.harvard.edu/abs/2011A%26A...528A.117P} {528, A117}

\bibitem[\protect\citeauthoryear{{Pakmor}, {Edelmann}, {R{\"o}pke}  \&
  {Hillebrandt}}{{Pakmor} et~al.}{2012a}]{pakmor+12aug}
{Pakmor} R.,  {Edelmann} P.,  {R{\"o}pke} F.~K.,   {Hillebrandt} W.,  2012a,
  \mn@doi [\mnras] {10.1111/j.1365-2966.2012.21383.x}, \href
  {http://adsabs.harvard.edu/abs/2012MNRAS.424.2222P} {424, 2222}

\bibitem[\protect\citeauthoryear{{Pakmor}, {Kromer}, {Taubenberger}, {Sim},
  {R{\"o}pke}  \& {Hillebrandt}}{{Pakmor} et~al.}{2012b}]{pakmor+12mar}
{Pakmor} R.,  {Kromer} M.,  {Taubenberger} S.,  {Sim} S.~A.,  {R{\"o}pke}
  F.~K.,   {Hillebrandt} W.,  2012b, \mn@doi [\apjl]
  {10.1088/2041-8205/747/1/L10}, \href
  {http://adsabs.harvard.edu/abs/2012ApJ...747L..10P} {747, L10}

\bibitem[\protect\citeauthoryear{{Pakmor}, {Kromer}, {Taubenberger}  \&
  {Springel}}{{Pakmor} et~al.}{2013}]{pakmor+2013}
{Pakmor} R.,  {Kromer} M.,  {Taubenberger} S.,   {Springel} V.,  2013, \mn@doi
  [\apjl] {10.1088/2041-8205/770/1/L8}, \href
  {http://adsabs.harvard.edu/abs/2013ApJ...770L...8P} {770, L8}

\bibitem[\protect\citeauthoryear{{Parrent} et~al.,}{{Parrent}
  et~al.}{2011}]{parrent+2011}
{Parrent} J.~T.,  et~al., 2011, \mn@doi [\apj] {10.1088/0004-637X/732/1/30},
  \href {http://adsabs.harvard.edu/abs/2011ApJ...732...30P} {732, 30}

\bibitem[\protect\citeauthoryear{{Parrent} et~al.,}{{Parrent}
  et~al.}{2012}]{parrent+2012}
{Parrent} J.~T.,  et~al., 2012, \mn@doi [\apjl] {10.1088/2041-8205/752/2/L26},
  \href {http://adsabs.harvard.edu/abs/2012ApJ...752L..26P} {752, L26}

\bibitem[\protect\citeauthoryear{{Paxton} et~al.,}{{Paxton}
  et~al.}{2015}]{paxton+2015}
{Paxton} B.,  et~al., 2015, \mn@doi [\apjs] {10.1088/0067-0049/220/1/15}, \href
  {http://adsabs.harvard.edu/abs/2015ApJS..220...15P} {220, 15}

\bibitem[\protect\citeauthoryear{{Plewa}}{{Plewa}}{2007}]{plewa+07}
{Plewa} T.,  2007, \mn@doi [\apj] {10.1086/511412}, \href
  {http://adsabs.harvard.edu/abs/2007ApJ...657..942P} {657, 942}

\bibitem[\protect\citeauthoryear{Plewa \& M\"uller}{Plewa \&
  M\"uller}{1999}]{PM99}
Plewa T.,  M\"uller E.,  1999, \aap, 342, 179

\bibitem[\protect\citeauthoryear{{Raskin}, {Scannapieco}, {Fryer},
  {Rockefeller}  \& {Timmes}}{{Raskin} et~al.}{2012}]{raskin+12}
{Raskin} C.,  {Scannapieco} E.,  {Fryer} C.,  {Rockefeller} G.,   {Timmes}
  F.~X.,  2012, \mn@doi [\apj] {10.1088/0004-637X/746/1/62}, \href
  {http://adsabs.harvard.edu/abs/2012ApJ...746...62R} {746, 62}

\bibitem[\protect\citeauthoryear{{Raskin}, {Kasen}, {Moll}, {Schwab}  \&
  {Woosley}}{{Raskin} et~al.}{2014}]{raskin+2014}
{Raskin} C.,  {Kasen} D.,  {Moll} R.,  {Schwab} J.,   {Woosley} S.,  2014,
  \mn@doi [\apj] {10.1088/0004-637X/788/1/75}, \href
  {http://adsabs.harvard.edu/abs/2014ApJ...788...75R} {788, 75}

\bibitem[\protect\citeauthoryear{{Ricker}}{{Ricker}}{2008}]{ricker2008}
{Ricker} P.~M.,  2008, \mn@doi [\apjs] {10.1086/526425}, \href
  {http://adsabs.harvard.edu/abs/2008ApJS..176..293R} {176, 293}

\bibitem[\protect\citeauthoryear{{Rosswog}}{{Rosswog}}{2015}]{rosswog+2015}
{Rosswog} S.,  2015, \mn@doi [Living Reviews in Computational Astrophysics]
  {10.1007/lrca-2015-1}, \href
  {http://adsabs.harvard.edu/abs/2015LRCA....1....1R} {1}

\bibitem[\protect\citeauthoryear{{Ruiz-Lapuente}}{{Ruiz-Lapuente}}{2014}]{ruiz-lapeunte+2014}
{Ruiz-Lapuente} P.,  2014, \mn@doi [\nar] {10.1016/j.newar.2014.08.002}, \href
  {http://adsabs.harvard.edu/abs/2014NewAR..62...15R} {62, 15}

\bibitem[\protect\citeauthoryear{{Saio} \& {Nomoto}}{{Saio} \&
  {Nomoto}}{1985}]{saio+85}
{Saio} H.,  {Nomoto} K.,  1985, \aap, \href
  {http://adsabs.harvard.edu/abs/1985A%26A...150L..21S} {150, L21}

\bibitem[\protect\citeauthoryear{{Sato}, {Nakasato}, {Tanikawa}, {Nomoto},
  {Maeda}  \& {Hachisu}}{{Sato} et~al.}{2015}]{sato+2015}
{Sato} Y.,  {Nakasato} N.,  {Tanikawa} A.,  {Nomoto} K.,  {Maeda} K.,
  {Hachisu} I.,  2015, \mn@doi [\apj] {10.1088/0004-637X/807/1/105}, \href
  {http://adsabs.harvard.edu/abs/2015ApJ...807..105S} {807, 105}

\bibitem[\protect\citeauthoryear{{Scalzo} et~al.,}{{Scalzo}
  et~al.}{2012}]{scalzo+2012}
{Scalzo} R.,  et~al., 2012, \mn@doi [\apj] {10.1088/0004-637X/757/1/12}, \href
  {http://adsabs.harvard.edu/abs/2012ApJ...757...12S} {757, 12}

\bibitem[\protect\citeauthoryear{{Sharpe}}{{Sharpe}}{2001}]{sharpe+01}
{Sharpe} G.~J.,  2001, \mn@doi [\mnras] {10.1046/j.1365-8711.2001.04119.x},
  \href {http://adsabs.harvard.edu/abs/2001MNRAS.322..614S} {322, 614}

\bibitem[\protect\citeauthoryear{{Shen} \& {Moore}}{{Shen} \&
  {Moore}}{2014}]{shen+2014}
{Shen} K.~J.,  {Moore} K.,  2014, \mn@doi [\apj] {10.1088/0004-637X/797/1/46},
  \href {http://adsabs.harvard.edu/abs/2014ApJ...797...46S} {797, 46}

\bibitem[\protect\citeauthoryear{{Sim}, {Fink}, {Kromer}, {R{\"o}pke}, {Ruiter}
   \& {Hillebrandt}}{{Sim} et~al.}{2012}]{sim+2012}
{Sim} S.~A.,  {Fink} M.,  {Kromer} M.,  {R{\"o}pke} F.~K.,  {Ruiter} A.~J.,
  {Hillebrandt} W.,  2012, \mn@doi [\mnras] {10.1111/j.1365-2966.2011.20162.x},
  \href {http://adsabs.harvard.edu/abs/2012MNRAS.420.3003S} {420, 3003}

\bibitem[\protect\citeauthoryear{{Tanikawa}, {Nakasato}, {Sato}, {Nomoto},
  {Maeda}  \& {Hachisu}}{{Tanikawa} et~al.}{2015}]{tanikawa+2015}
{Tanikawa} A.,  {Nakasato} N.,  {Sato} Y.,  {Nomoto} K.,  {Maeda} K.,
  {Hachisu} I.,  2015, \mn@doi [\apj] {10.1088/0004-637X/807/1/40}, \href
  {http://adsabs.harvard.edu/abs/2015ApJ...807...40T} {807, 40}

\bibitem[\protect\citeauthoryear{{Timmes}}{{Timmes}}{1999}]{timmes99}
{Timmes} F.~X.,  1999, \apjs, 124, 241

\bibitem[\protect\citeauthoryear{Timmes}{Timmes}{2016}]{ftimmeswebsite}
Timmes F.,  2016, {http://cococubed.asu.edu/code\_pages/burn\_helium.shtml}

\bibitem[\protect\citeauthoryear{{Timmes} \& {Swesty}}{{Timmes} \&
  {Swesty}}{2000}]{timmes+2000}
{Timmes} F.~X.,  {Swesty} F.~D.,  2000, \mn@doi [\apjs] {10.1086/313304}, \href
  {http://adsabs.harvard.edu/abs/2000ApJS..126..501T} {126, 501}

\bibitem[\protect\citeauthoryear{{Toonen}, {Nelemans}  \& {Portegies
  Zwart}}{{Toonen} et~al.}{2012}]{toonen+12}
{Toonen} S.,  {Nelemans} G.,   {Portegies Zwart} S.,  2012, \mn@doi [\aap]
  {10.1051/0004-6361/201218966}, \href
  {http://adsabs.harvard.edu/abs/2012A%26A...546A..70T} {546, A70}

\bibitem[\protect\citeauthoryear{{Travaglio}, {Kifonidis}  \&
  {M{\"u}ller}}{{Travaglio} et~al.}{2004}]{travaglio+2004}
{Travaglio} C.,  {Kifonidis} K.,   {M{\"u}ller} E.,  2004, \mn@doi [\nar]
  {10.1016/j.newar.2003.11.046}, \href
  {http://adsabs.harvard.edu/abs/2004NewAR..48...25T} {48, 25}

\bibitem[\protect\citeauthoryear{{Turk}, {Smith}, {Oishi}, {Skory}, {Skillman},
  {Abel}  \& {Norman}}{{Turk} et~al.}{2011}]{turk+2011}
{Turk} M.~J.,  {Smith} B.~D.,  {Oishi} J.~S.,  {Skory} S.,  {Skillman} S.~W.,
  {Abel} T.,   {Norman} M.~L.,  2011, \mn@doi [\apjs]
  {10.1088/0067-0049/192/1/9}, \href
  {http://adsabs.harvard.edu/abs/2011ApJS..192....9T} {192, 9}

\bibitem[\protect\citeauthoryear{{Weaver}, {Zimmerman}  \& {Woosley}}{{Weaver}
  et~al.}{1978}]{weaver+78}
{Weaver} T.~A.,  {Zimmerman} G.~B.,   {Woosley} S.~E.,  1978, \mn@doi [\apj]
  {10.1086/156569}, \href {http://adsabs.harvard.edu/abs/1978ApJ...225.1021W}
  {225, 1021}

\bibitem[\protect\citeauthoryear{{Webbink}}{{Webbink}}{1984}]{webbink+84}
{Webbink} R.~F.,  1984, \mn@doi [\apj] {10.1086/161701}, \href
  {http://adsabs.harvard.edu/abs/1984ApJ...277..355W} {277, 355}

\bibitem[\protect\citeauthoryear{{Whelan} \& {Iben}}{{Whelan} \&
  {Iben}}{1973}]{whelan+73}
{Whelan} J.,  {Iben} Jr. I.,  1973, \mn@doi [\apj] {10.1086/152565}, \href
  {http://adsabs.harvard.edu/abs/1973ApJ...186.1007W} {186, 1007}

\bibitem[\protect\citeauthoryear{{Woosley}, {Taam}  \& {Weaver}}{{Woosley}
  et~al.}{1986}]{Woosley+86}
{Woosley} S.~E.,  {Taam} R.~E.,   {Weaver} T.~A.,  1986, \apj, 301, 601

\bibitem[\protect\citeauthoryear{{Woosley}, {Wunsch}  \& {Kuhlen}}{{Woosley}
  et~al.}{2004}]{woosley+04}
{Woosley} S.~E.,  {Wunsch} S.,   {Kuhlen} M.,  2004, \mn@doi [\apj]
  {10.1086/383530}, \href {http://adsabs.harvard.edu/abs/2004ApJ...607..921W}
  {607, 921}

\bibitem[\protect\citeauthoryear{{Yoon}, {Podsiadlowski}  \& {Rosswog}}{{Yoon}
  et~al.}{2007}]{yoon+07}
{Yoon} S.-C.,  {Podsiadlowski} P.,   {Rosswog} S.,  2007, \mn@doi [\mnras]
  {10.1111/j.1365-2966.2007.12161.x}, \href
  {http://adsabs.harvard.edu/abs/2007MNRAS.380..933Y} {380, 933}

\bibitem[\protect\citeauthoryear{{de Val-Borro} et~al.,}{{de Val-Borro}
  et~al.}{2006}]{valborro+2006}
{de Val-Borro} M.,  et~al., 2006, \mn@doi [\mnras]
  {10.1111/j.1365-2966.2006.10488.x}, \href
  {http://adsabs.harvard.edu/abs/2006MNRAS.370..529D} {370, 529}

\bibitem[\protect\citeauthoryear{{van Kerkwijk}, {Chang}  \& {Justham}}{{van
  Kerkwijk} et~al.}{2010}]{vanKerkwijk+2010}
{van Kerkwijk} M.~H.,  {Chang} P.,   {Justham} S.,  2010, \mn@doi [\apjl]
  {10.1088/2041-8205/722/2/L157}, \href
  {http://adsabs.harvard.edu/abs/2010ApJ...722L.157V} {722, L157}

\makeatother
\end{thebibliography}
